%% file: paper.tex
\documentclass{vldb}

\usepackage{xcolor}
\usepackage{booktabs}
\usepackage{pgfplots}
\usepackage{tikz}
\usepackage{multirow}
\usepackage{graphicx}
\usepackage[final]{microtype}
\usepackage[sort,nocompress]{cite}
\usepackage{amsmath}
\usepackage{siunitx}
\usepackage{array}
\usepackage[normalem]{ulem}
\usepackage{xcolor}
\usepackage{url}
\usepackage{adjustbox}
\usepackage{subfig}

\usetikzlibrary{arrows, positioning, automata}
\usetikzlibrary{fit,chains, calc}
\usetikzlibrary{shapes.geometric}
\usetikzlibrary{arrows.meta}

\usepackage{macrosreport}

\includecomment{tr} \excludecomment{paper}

\vldbTitle{Materializing Knowledge Bases via Trigger Graphs}
\vldbAuthors{}
\vldbDOI{https://doi.org/10.14778/xxxxxxx.xxxxxxx}
\vldbVolume{12}
\vldbNumber{xxx}
\vldbYear{2020}

\begin{document}

\title{Materializing Knowledge Bases via Trigger Graphs (Technical Report)}
\numberofauthors{1}
\author{Efthymia Tsamoura$^{*}$, David Carral$^{\dagger}$, Enrico Malizia$^{\ddagger}$, Jacopo Urbani$^{\diamond}$\\
\affaddr{$^{*}$Samsung AI Research, United Kingdom; $^{\dagger}$TU Dresden, Germany;}
\\
\affaddr{$^{\ddagger}$ University of Bologna, Italy; $^{\diamond}$Vrije
Universiteit Amsterdam, The Netherlands}
}

\maketitle

\input{sections/abstract}
\input{sections/introduction}

\input{sections/motivation}

\input{sections/preliminaries}

\input{sections/trigger-graphs}
\input{sections/computing-trigger-graphs}

\input{sections/linear-minimization}
\input{sections/online}
\input{sections/evaluation}
\input{sections/related}

\input{sections/conclusions}

\bibliographystyle{abbrv}
\bibliography{paper}
\newpage
\onecolumn

\begin{tr}
	\onecolumn

	\appendix

	\input{appendix/experimental-results}

	\newpage
\section{Additional definitions}
	\input{appendix/definitions}
\section{Proofs for results in Section 4}
	\input{appendix/fulltg}

	\input{appendix/classes}
	\input{appendix/ftgbdd}
	\input{appendix/bddtdbfus}
	\input{appendix/ftgundecidability}
	\input{appendix/existence}
\section{Proofs for results in Section 5}
	\input{appendix/tglinear}
	\input{appendix/tglinear-complexity}
	\input{appendix/minlinear}
\section{Proofs for results in Section 5.1}
	\input{appendix/EG-rewriting}
	\input{appendix/minDatalog}

	\input{appendix/complexity_minDatalog}
	\input{appendix/TGmat}
	\input{appendix/examples}

\end{tr}
\end{document}

%% file: sections/abstract.tex
\begin{abstract}


\smallchg{%
The \emph{chase} is a well-established family of algorithms used to materialize
Knowledge Bases (KBs), like Knowledge Graphs (KGs), to tackle important tasks
like query answering under dependencies or data cleaning. A general problem of
chase algorithms is that they might perform redundant computations.}
To counter this problem, we introduce the notion of \emph{Trigger Graphs} (TGs),
which guide the execution of the rules avoiding redundant computations. We
present the results of an extensive theoretical and empirical study that seeks
to answer when and how TGs can be computed and what are the benefits of TGs when
applied over real-world KBs. Our results include introducing algorithms that
compute (minimal) TGs. \smallchg{We implemented our approach in a new engine,
    and our experiments show that it can be significantly more efficient than
the chase enabling us to materialize KBs with 17B facts in less than 40 min on
commodity machines.}

\end{abstract}

%% file: sections/introduction.tex
\section{Introduction}

\leanparagraph{Motivation} Knowledge Bases (KBs) are becoming increasingly
important with many industrial key players investing on this technology. For
example, Knowledge Graphs (KGs)~\cite{kg} have emerged as the main vehicle for
representing factual knowledge on the Web and enjoy a widespread
adoption~\cite{noy_industry-scale_2019}.
\smallchg{Moreover, several tech giants are building KGs to support their core
    business. For instance, the KG developed at Microsoft contains information about the world
    and supports question answering, while, at Google, KGs are used to help Google
    products respond more appropriately to user requests by mapping them to
    concepts in the KG. 
    The use
    of KBs and KGs in such scenarios is not restricted only to database-like
    analytics or query answering: KBs play also a central role in
    neural-symbolic systems for efficient learning and explainable AI
\cite{DBLP:books/daglib/0007534,DBLP:conf/semweb/KruitBU19}.  }

A KB can be viewed as a classical database $B$ with factual knowledge and a set
of logical rules $P$, called \emph{program}, \smallchg{allowing} the derivation of
additional knowledge. \smallchg{One class of rules that is of particular
interest both to academia and to industry is Datalog \cite{alice}.  Datalog
is a recursive language with declarative semantics that allows users to
succinctly write recursive graph queries. Beyond expressing graph queries,
e.g., reachability, Datalog allows richer fixed-point graph analytics via
aggregate functions. LogicBlox and LinkedIn used Datalog to develop
high-performance applications, or to compute analytics over its KG
\cite{LogicBlox,Datalography}. Google developed their own Datalog engine
called Yedalog~\cite{yedalog}.  Other industrial users include Facebook,
BP \cite{vadalog} and Samsung \cite{recom}.

\emph{Materializing} a KB ${(P,B)}$ is the process of deriving all the
facts that logically follow when reasoning over the database $B$ using the rules
in $P$.  Materialization is a core operation in KB management. An obvious use
is that of caching the derived knowledge. A second use is that of
\emph{goal-driven query answering}, i.e., deriving the knowledge specific to a
given query \emph{only}, using database techniques such as magic sets and
subsumptive tabling
\cite{DBLP:conf/pods/BancilhonMSU86,DBLP:journals/jlp/BeeriR91,10.1145/1989323.1989393,AAAI1816927}.
Beyond knowledge exploration, other applications of materialization are
data wrangling~\cite{konstantinou_vada_2017},
entity resolution~\cite{kruit2020tab2know}, data exchange~\cite{Fagin2005a}
and query answering over OWL~\cite{owl} and RDFS~\cite{rdfs} ontologies.
Finally, materialization has
been also used in probabilistic KBs \cite{DBLP:conf/aaai/TsamouraGK20}.

\leanparagraph{Problem} The increasing sizes of modern KBs
\cite{noy_industry-scale_2019}, 
 and 
 the fact that
materialization is not a one-off operation when used for goal-driven query
answering, make improving the materialization
performance critical. The \emph{chase}, which was introduced in {1979 by Maier et
al.~\cite{10.1145/320107.320115}}, has been the most popular materialization
technique and has been adopted by several commercial and open source engines such as
VLog~\cite{vlog},  RDFox~\cite{rdfox} and Vadalog~\cite{vadalog}.


To improve the performance of materialization, different approaches have focused
on different inefficiency aspects. One approach is to reduce the number of facts
added in the KB. This is the take of some of the chase variants proposed by the
database and AI communities~\cite{benchmarking-chase,onet,k-boundedness}. A
second approach is to parallelize the computation. For example, RDFox proposes a
parallelization technique for Datalog rules \cite{rdfox}, while WebPIE
\cite{webpie} and Inferray \cite{Inferray} propose parallelization techniques
for fixed RDFS rules. Orthogonal to those approaches are those employing
compression and columnar storage layouts to reduce memory
consumption~\cite{vlog,hu_datalog_2019}.

In this paper, we focus on a different aspect: that of avoiding
redundant computations. Redundant computations is a problem that concerns all
chase variants and has multiple causes. A first cause is the
derivation of facts that either have been derived in previous rounds, or are
logically redundant, i.e., they can be ignored without compromising query
answering. The above issue has been partially addressed in Datalog with the
well-known semina\"ive evaluation (SNE)~\cite{alice}. SNE restricts the
execution of the rules over at least one new fact.  However, it cannot block the
derivation of the same or logically redundant facts by different rules.
A second cause of redundant computations relates to the execution of
the rules: when executing a rule, the chase may consider facts that cannot lead
to any derivations.}

\leanparagraph{Our approach} To reduce the amount of redundant computations, we introduce the
notion of \emph{Trigger Graphs (TGs)}. A TG is an acyclic directed graph that
captures all the operations that should be performed to materialize a KB
(${P,B}$). Each node in a TG is associated with a rule from $P$ and \smallchg{with} a set of
facts, while the edges specify the facts over which we execute each rule.

Intuitively, a TG can be viewed as a blueprint for reasoning over the KB.  As
such, we can use it to ``guide'' a reasoning procedure without resorting to an
exhaustive execution of the rules, as it is done with the chase.  In particular,
our approach consists of traversing the TG, executing the rule $r$ associated
with a node $v$ over the union of the facts associated with the parent nodes of
$v$ and storing the derived facts \smallchg{``}inside\smallchg{''} $v$. After the traversal is complete,
then the materialization of the KB is simply the union of the facts in all the
nodes.

\smallchg{
TG-guided materialization addresses \emph{at the same time all} causes of inefficiencies
described above. In particular, TGs block the derivation of the same or
logically redundant facts that cannot be blocked by SNE. This is achieved by effectively partitioning
into smaller sub-instances the facts currently in the KB.
This partitioning also enables us to reduce the cost of executing the rules.

Furthermore, in specific cases,
TGs allow us reasoning via either completely avoiding certain steps involved when executing rules,
or performing them at the end and collectively for all
rules. Our experiments show that we get
good runtime improvements with both alternatives.
}

\smallchg{
\leanparagraph{Contributions} We propose techniques for computing
both instance-independent and instance-dependent TGs.
The former TGs are computed exclusively based on the rules of the KB
and allow us to reason over \emph{any} possible instance of the KB
making them particularly useful when the database changes frequently.
In contrast, instance-dependent TGs are computed based both on the rules and the data of the KB
and, thus, support reasoning over the given KB \emph{only}.}
We show that not every program admits a finite instance-independent TG.
We define a special class, called \Ftg, including all programs that admit a finite
instance-independent TG and explore its relationship with other
known classes.

As a second contribution, we propose algorithms {to} compute and minimize
(instance-independent) TGs for linear programs: a class of programs relevant in practice.
%
A program $P$ not admitting a finite \emph{instance-independent} TG may still
admit a finite \emph{instance-dependent} TG.

As a third contribution, we show that all programs that
admit a finite universal model also admit a finite \emph{instance-dependent} TG.
We use this finding to propose a TG-guided materialization technique that
supports \emph{any} such program (not necessarily in \Ftg). The technique
works by interleaving the reasoning process with the computation of the TG, and it
reduces the number of redundant computations via query containment and via a novel
TG-based rule execution strategy.

\smallchg{
We implemented our approach in a new reasoner, called \sys{}, and
compared its performance versus multiple state-of-the-art chase and RDFS engines
including RDFox, VLog, WebPIE \cite{webpie} and Inferray \cite{Inferray}, using well-established
benchmarks, e.g., \chaseBench \cite{benchmarking-chase}.
Our evaluation shows that GLog outperforms all its competitors in all
benchmarks.
Moreover, in our largest experiment, \sys{} was able to materialize a KB with
17B facts in 37 minutes on commodity hardware.
}

\leanparagraph{Summary} We make the following contributions:
\begin{compactitem}

\item (\emph{New idea}) We propose a new reasoning technique based on traversing
    acyclic graphs, called TGs, to tackle multiple sources of
    inefficiency of the chase;

\item (\emph{New theoretical contribution}) We study the class of programs
    admit{ting} finite instance-independent TGs and its relationship with other
    known classes;

\item (\emph{New algorithms}) We propose techniques for computing minimal
    instance-independent TGs for linear programs\smallchg{, and} techniques for computing
    minimal instance-dependent TGs for Datalog programs;

\item (\emph{New system}) We introduce a new reasoner, \sys{}, which has
competitive performance, often superior to the state-of-the-art, and has
good scalability.  \end{compactitem}

Supplementary material with all proofs, code and evaluation data is in
\url{https://bitbucket.org/tsamoura/trigger-graphs/src/master/}.

%% file: sections/motivation.tex
\section{Motivating Example}\label{section:motivation}

\begin{figure}[t] \centering
    \includegraphics[width=\columnwidth]{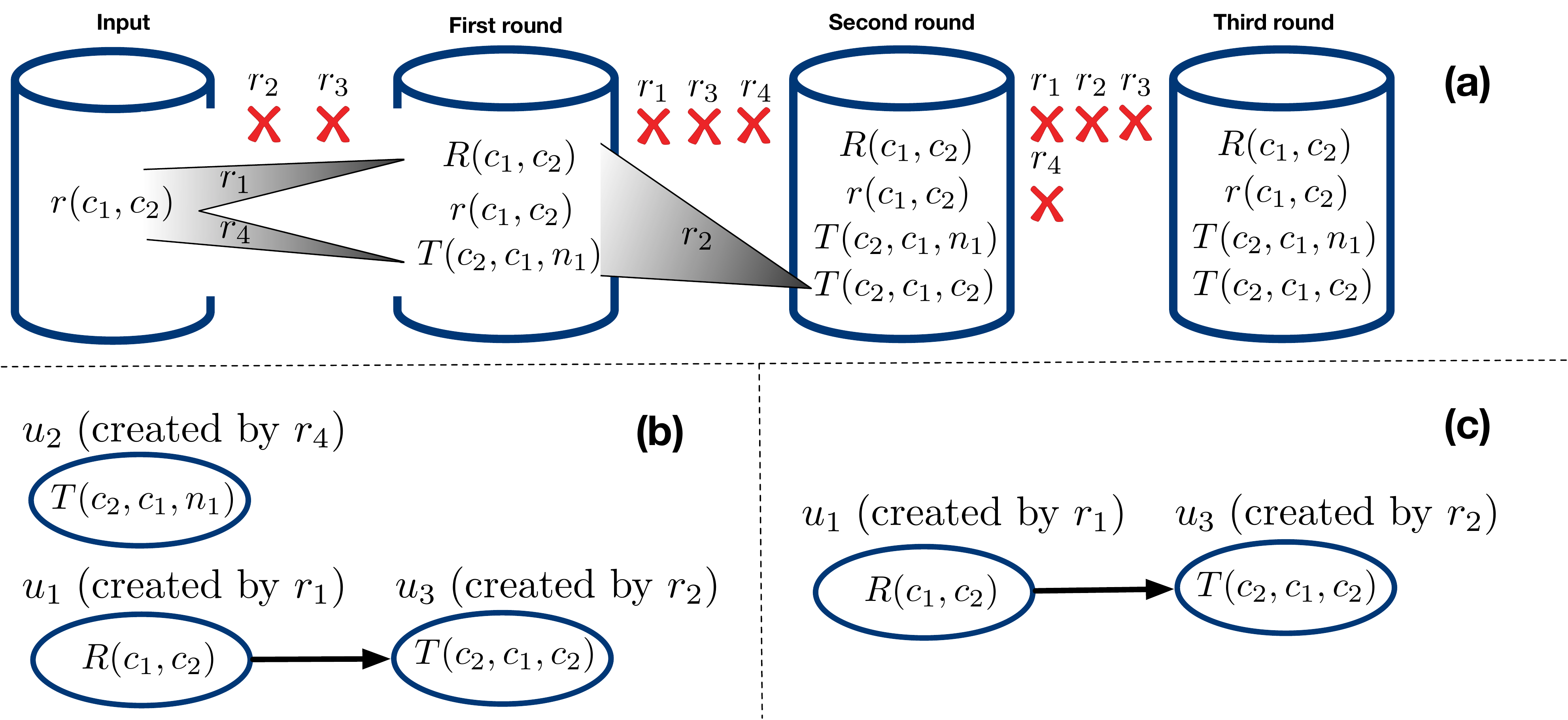}
    \caption{\smallchg{(a)
        Chase execution for Example~\ref{example:running:demo}, (b) the TG
        $G_1$, (c) the TG $G_2$. In (b) and (c), the facts shown inside the nodes are
the results of reasoning over $B$ using the TG.}}\label{figure:trigger-graphs}
\end{figure}

\smallchg{
We start our discussion with a simple example to describe how the chase works,
its inefficiencies, and how they can be overcome with TGs. For the moment, we
give only an intuitive description of some key concepts to aid the understanding
of the main ideas. In the following sections, we will provide a formal
description.

The chase works in rounds during which it executes the rules over the facts that
are currently in the KB. In most chase variants, the execution of a rule
involves three steps: retrieving all the facts that instantiate the premise of
the rule, then, checking whether the facts to be derived logically hold in the
KB and finally, adding them to the KB if they do.

\begin{example} \label{example:running:demo}

Consider the KB $B=\{r(c_1,c_2)\}$ with a single fact and the program ${P_1 =
\{r_1,r_2,r_3,r_4\}}$:
\begin{align} r(X,Y) &\rightarrow R(X,Y)    \subtag{$r_1$} \\ R(X,Y)
&\rightarrow T(Y,X,Y)  \subtag{$r_2$} \\ T(Y,X,Y) &\rightarrow R(X,Y)
\subtag{$r_3$} \\ r(X,Y) &\rightarrow \exists Z. T(Y,X,Z)   \subtag{$r_4$}
\end{align}
Figure~\ref{figure:trigger-graphs} (a) depicts the rounds of the chase with such
an input. In the first round, the only rules that can derive facts are $r_1$
and $r_4$. Rule $r_1$ derives the fact ${R(c_1,c_2)}$. Since this fact is not
in the KB, the chase adds it to the KB. Let us now focus on $r_4$. Notice that
variable $Z$ in $r_4$ does not occur in the premise of $r_4$. The chase deals
with such variables by introducing fresh \emph{null (values)}.  Nulls can be
seen as ``placeholders'' for objects that are not known. In our case, $r_4$
derives the fact ${T(c_2,c_1,\lnu_1)}$, where $\lnu_1$ is a null, and the chase
adds it to the KB.

The chase then continues to the second round where rules are executed over
$B'=B\cup \{ R(c_1,c_2), T(c_2,c_1,\lnu_1)\}$. The execution of $r_2$ derives
the fact ${T(c_2,c_1,c_2)}$, which is added to the KB, yielding $B''=B'\cup
\{T(c_2,c_1,c_2)\}$.
Finally, the chase proceeds to the third round where only rule $r_3$ derives
${R(c_1,c_2)}$ from $B''$. However, since this fact is already in
$B''$, the chase stops.  \end{example}

The above steps expose two inefficiencies of the chase. The first is that of
incurring in the cost of deriving the same or logically redundant facts.

\begin{example} Let us return back to Example~\ref{example:running:demo}. The
    chase pays the cost of executing $r_3$ despite that $r_3$'s execution
    always derives facts derived in previous rounds.  This is due to
    the cyclic dependency between rules $r_2$ and $r_3$: $r_2$ derives
    $T$-facts by flipping the arguments of the $R$-facts, while $r_3$ derives
    $R$-facts by flipping the arguments of the $T$-facts.  Despite that the SNE
    effectively blocks the execution of $r_1$ and $r_2$ in the third chase
    round, it cannot block the execution of $r_3$ in the third chase round,
    since ${T(c_2,c_1,c_2)}$ was derived in the second round.

	Now, consider the fact ${T(c_2,c_1,\lnu_1)}$. This fact is logically
	redundant, because it provides no extra information over
	${T(c_2,c_1,c_2)}$, derived by $r_2$. Despite being logically redundant,
	the chase pays the cost of deriving it.
\end{example}

The second inefficiency that is exposed is that of suboptimally executing the
rules themselves: when computing the facts instantiating the premise of a rule,
the chase considers all facts currently in the KB even the ones that cannot
instantiate the premise of the rule.

\begin{example}	 Continuing with Example~\ref{example:running:demo}, consider
    the execution of $r_3$ in the second round of the chase.  No fact derived by
    $r_4$ can instantiate the premise of $r_3$, since the premise of $r_3$
    requires the first and the third arguments of the $T$-facts to be the same.
Hence, the cost paid for executing $r_3$ over those facts is unnecessary.
\end{example}

The root of these inefficiencies is that the chase, in each round, considers the
entire KB as a source for potential derivations, with only SNE as means to avoid
some redundant derivations. If we were able to ``guide'' the execution of the
rules in a more clever way, then we can avoid the inefficiencies stated above.

For instance, consider an alternative execution strategy where $r_2$ is executed
\emph{only} over the derivations of $r_1$, while $r_3$ and $r_4$ are not
executed at all. This strategy would not face any of the inefficiencies
highlighted above, and it can be defined with a graph like the one in
Figure~\ref{figure:trigger-graphs} (c). Informally, a \emph{Trigger Graph} (TG)
is precisely such a graph-based blueprint to compute the materialization. In
the remaining, we will first provide a formal definition of TGs and study their
properties. Then, we will show that in some cases we can build a TG that is
optimal for any possible set of facts given as input. In other cases, we can
still build TGs incrementally.  Such TGs allow to avoid redundant computations
that will occur with the chase but only with the given input.  }

%% file: sections/preliminaries.tex
\section{Preliminaries}
\label{section:preliminaries}

Let \Consts, \Nulls, \Vars, and \Preds be mutually disjoint, (countably infinite) sets of \emph{constants}, \emph{nulls}, \emph{variables}, and \emph{predicates}, respectively.
Each predicate $p$ is associated with a non-negative integer ${\Ar{p} \geq 0}$, called the \emph{arity} of $p$.
Let \EP and \IP be disjoint subsets of \Preds of \emph{intensional} and \emph{extensional predicates}, respectively.
A \emph{term} is a constant, a null, or a variable.
A term is ground if it is either a constant or a null.
An \emph{atom} $\At$ has the form ${p(t_1,\dots,t_n)}$, where $p$ is an $n$-ary predicate,
and ${t_1,\dots,t_n}$ are terms.
An atom \At is extensional (resp., intensional), if the predicate of $\At$ is in $\EP$ (resp., $\IP$).
A \emph{fact} is an atom of ground terms.
A \emph{base fact} is an atom of constants whose predicate is extensional.
An \emph{instance} $\I$ is a set of facts (possibly comprising null terms).
A \emph{base instance} $\BI$ is a set of base facts.

A \emph{rule} is a first-order formula of the form
\begin{equation}\label{rule}
    \forall \vec{X} \forall \vec{Y}
    \bigwedge \nolimits_{i=1}^n P_i(\vec{X}_i,\vec{Y}_i) \to
    \exists \vec{Z} P(\vec{Y},\vec{Z}),
\end{equation}
where, $P$ is an intensional predicate and for all $1\leq i\leq n$, $\vec{X}_i\subseteq\vec{X}$ and $\vec{Y}_1\subseteq\vec{Y}$ ($\vec{X}_i$ and $\vec{Y}_i$ might be empty).
We assume w.l.o.g.\ that
the body of a rule includes only extensional predicates or intensional predicates.
We will denote extensional predicates with lowercase letters, while intensional predicates with uppercase letters.
Quantifiers are commonly omitted.
The left-hand and the right-hand side of a rule \ERule are its \emph{body} and
\emph{head}, respectively, and are denoted by $\body{r}$ and $\head{r}$.
A rule is \emph{Datalog} if it has no existentially quantified variables,
\emph{extensional} if $\body{r}$ includes only extensional atoms, and \emph{linear}
if it has a single atom in its body.

A \emph{program} is a set of rules.
A \emph{knowledge base} (KB) is a pair ${( \P, \BI )}$ with \P a program and \BI a base instance.

Symbol $\models$ denotes logical entailment, where sets of atoms and rules are viewed as first-order theories.
Symbol $\equiv$ denotes logical equivalence, i.e.,\ logical entailment in both directions.

\smallchg{%
A \emph{term mapping} $\sigma$ is a (possibly partial) mapping of terms to terms; we write
${\sigma = \{ t_1 \mapsto s_1, \dots, t_n \mapsto s_n \}}$ to denote that
${\sigma(t_i) = s_i}$ for ${1 \leq i \leq n}$. Let $\alpha$ be a term, an atom, a
conjunction of atoms, or a set of atoms. Then $\sigma(\alpha)$ is obtained by
replacing each occurrence of a term $t$ in $\alpha$ that also occurs in the
domain of $\sigma$ with $\sigma(t)$ (i.e., terms outside the domain of $\sigma$
remain unchanged).
A \emph{substitution} is a term mapping whose domain contains only variables and whose range contains only ground terms.
For two sets, or conjunctions, of atoms $\mathcal{A}_1$ and $\mathcal{A}_2$, a term mapping $\sigma$ from
the terms occurring in $\mathcal{A}_1$ to the terms occurring in $\mathcal{A}_2$}
is said to be a \emph{homomorphism} from $\mathcal{A}_1$ to $\mathcal{A}_2$ if the following hold:
(i) $\sigma$ maps each constant in its domain to itself,
(ii) $\sigma$ maps each null in its domain to ${\Consts \cup \Nulls}$ and
(iii) for each atom $A\in \mathcal{A}_1$, ${\sigma(A) \in \mathcal{A}_2}$.
We denote a homomorphism $\sigma$ from $\mathcal{A}_1$ into $\mathcal{A}_2$ by
$\sigma: \mathcal{A}_1 \to \mathcal{A}_2$.

It is known that, for two sets of facts $\mathcal{A}_1$ and $\mathcal{A}_2$,
there exists a homomorphism from $\mathcal{A}_1$ into $\mathcal{A}_2$
iff ${\mathcal{A}_2 \models \mathcal{A}_1}$ (and hence, there exists a homomorphism in both ways
iff $\mathcal{A}_1\equiv\mathcal{A}_2$).
When $\mathcal{A}_1$ and $\mathcal{A}_2$ are null-free instances,
${\mathcal{A}_2 \models \mathcal{A}_1}$ iff ${\mathcal{A}_1 \subseteq \mathcal{A}_2}$
and ${\mathcal{A}_2 \equiv \mathcal{A}_1}$ iff ${\mathcal{A}_1 = \mathcal{A}_2}$.

\smallchg{%
For a set of two or more atoms ${\mathcal{A} = \{A_1,\dots,A_n\}}$
a \emph{most general unifier} (MGU) $\mu$ for $\mathcal{A}$ is a substitution so that: (i) ${\mu(A_1) = \dots = \mu(A_n)}$; and (ii)
for each other substitution $\sigma$ for which ${\sigma(A_1) = \dots = \sigma(A_n)}$, there exists a $\sigma'$ such that ${\sigma = \sigma' \circ \mu}$ \cite{10.5555/320343}.
}


Consider a rule $r$ of the form~\eqref{rule} and an instance $I$.
A \emph{trigger} for $r$ in $I$ is a homomorphism from
the body of $r$ into $I$.
\smallchg{We denote by $h_s$ the extension of a trigger $h$ mapping each
${Z \in \vec{Z}}$ into a unique fresh null.}
A rule $r$ \emph{holds} or is \emph{satisfied} in an instance $I$, if \bigchg{for each trigger $h$
for $r$ in $I$, there exists an extension $h'$ of $h$ to a homomorphism from the head of $r$ into $I$}.
A \emph{model} of a KB ${(P,B)}$ is a set ${I \supseteq B}$, such that each ${r \in P}$ holds in $I$.
A KB may admit infinitely many different models.
A model $M$ is \emph{universal},
if there exists a homomorphism from $M$ into every other model of ${(P,B)}$.
A program $P$ is \emph{Finite Expansion Set} (\Fes), if for each base instance $B$, ${(P,B)}$ admits a finite universal model.

A \emph{conjunctive query} (CQ) is a formula of the form ${Q(X_1,\dots,X_n) \leftarrow \bigwedge_{i = 1}^m A_i}$,
where $Q$ is a fresh predicate not occurring in $P$,
$\At_i$ are null-free atoms and each $X_j$ occurs in some $A_i$ atom.
We usually refer to a CQ by its head predicate.
We refer to the left-hand and the right-hand side of the formula as the \emph{head} and the \emph{body} of the query, respectively.
A CQ is \emph{atomic} if its body consists of a single atom.
A Boolean CQ (BCQ) is a CQ whose head predicate has no arguments.
A substitution $\sigma$ is an \emph{answer} to $Q$ on an instance $I$ if the domain of $\sigma$
is precisely its head variables, and if $\sigma$ can be extended to a
homomorphism from ${\bigwedge_i A_i}$ into $I$.
We often identify $\sigma$ with the $n$-tuple ${(\sigma(X_1), \ldots, \sigma(X_n))}$.
The \emph{output} of $Q$ on $I$
is the set $Q(I)$ of all answers to $Q$ on $I$.
The answer to a BCQ $Q$ on an instance $I$ is true, denoted as ${I \models Q}$,
if there exists a homomorphism from $\bigwedge_{i = 1} A_i$ into $I$.
The answer to a BCQ $Q$ on a KB $(P,B)$ is true, denoted as ${(P,B) \models Q}$,
if ${M \models Q}$ holds, for each model $M$ of $(P,B)$.
Finally, a CQ $Q_1$ is \emph{contained} in a CQ $Q_2$, denoted as
${Q_1 \subseteq Q_2}$, if for each instance $I$,
each answer to $Q_1$ on $I$ is in the answers to $Q_2$ on $I$ \cite{Chandra:1977:OIC:800105.803397}.

\smallchg{%
The \emph{chase} refers to a family of techniques for repairing a base instance $B$ relative to a set of rules $P$
so that the result satisfies the rules in $P$ and contains all base facts from $B$.
In particular, the result is a universal model of $(P,B)$, which we can use for query answering \cite{Fagin2005a}.
By ``chase'' we refer both to the procedure and its output.

The chase works in rounds during which it executes one or more rules from the KB.
The result of each round ${i \geq 0}$ is a new instance $I^i$ (with ${I^0 = B}$),
which includes the facts of all previous instances plus the newly derived facts.
The execution of a rule in the $i$-th chase round, involves computing all triggers from the body of $r$ into $I^{i-1}$, then
(potentially) checking whether the facts to be derived satisfy certain criteria in the KB and finally,
adding to the KB or discarding the derived facts.
Different chase variants employ different criteria for deciding
whether a fact should be added to the KB or whether to stop or continue the reasoning process \cite{benchmarking-chase,onet}.
For example, the restricted chase (adopted by VLog and RDFox)
adds a fact if there exists no homomorphism from this fact into the KB and terminates when no new fact is added.
The warded chase (adopted by Vadalog) replaces homomorphism checks by isomorphism ones \cite{vadalog} and terminates, again,
when no new fact is added.
The equivalent chase omits any checks and terminates when there is a round $i$ which produces an instance
that is logically equivalent to the instance produced in the ${(i-1)}$-th round \cite{k-boundedness}.
Notice that when a KB includes only Datalog rules all chase variants behave the same:
a fact is added when it has not been previously derived and the chase stops when no new fact is added to the KB.

Not all chase variants terminate even when the KB admits a finite universal model \cite{k-boundedness}.
The core chase \cite{chaserevisited} and the equivalent one do offer such guarantees.

For a chase variant, we use $\Step{i}{\K}$ or $\Step{i}{P,B}$ to denote the instance
computed during the $i$-th chase round and ${\Chase{P,B}}$ to denote the (possibly infinite) result of the chase.
Furthermore, we define the \emph{chase graph} ${\chaseGraph(P,B)}$ for a KB ${(P,B)}$ as the edge-labeled directed acyclic graph
having as nodes the facts in ${\Chase{P,B}}$ and having an edge
from a node $f_1$ to $f_2$ labeled with rule ${r \in P}$ if $f_2$ is obtained from $f_1$ and possibly from other facts
by executing ${r}$.
}

%% file: sections/trigger-graphs.tex
\section{Trigger Graphs}\label{section:trigger-graph}


In this section, we formally define Trigger Graphs (TGs) and
study the class of programs admitting finite
\emph{instance-independent} TGs.
First, we introduce the notion of \emph{Execution Graphs} (EGs).
Intuitively, an EG for a program is a digraph stating a ``plan'' of rule execution to reason via the program.
In its general definition, an EG is not required to characterize a plan of reasoning guaranteeing completeness.
Particular EGs, defined later, will also satisfy this property.


\begin{definition}
\label{definition:execution-graph}
	An \emph{execution graph} (EG) for a program \P is an acyclic, node- and edge-labelled digraph ${\EG = ( V, E , \rulname, \ell)}$,
    where $V$ and $E$ are the \smallchg{graph nodes and edges sets}, respectively, and $\rulname$ and $\ell$ are the node- and edge-labelling functions.
	\smallchg{Each node $v$
	(i) is labelled with some rule, denoted by $\rul{v}$, from $P$; and
	(ii) if the $j$-th predicate in the body of $\rul{v}$ equals the head
    predicate of $\rul{u}$ for some node $u$, then there is an edge labelled $j$
from node $u$ to node~$v$, denoted by $u \rightarrow_j v$.}
\end{definition}

Figures~\ref{figure:trigger-graphs}(b) and \ref{figure:trigger-graphs}(c)
show two EGs for $P_1$ from Example~\ref{example:running:demo}.
Next to each node is the associated rule.
Later we show that both EGs are also TGs for $P_1$.

Since the nodes of an execution graph are associated with rules of a program,
when, in the following, we refer to the head and the body of a node $v$, we actually mean the head and the body of $\rul{v}$.
Observe that, by definition, nodes associated with extensional rules do not have entering edges,
and nodes $v$ associated with an intensional rule have at most \emph{one}
incoming edge associated with the $j$-th predicate of the body of $v$, i.e., there is at most one node $u$ such that ${u \rightarrow_j v}$.
The latter might seem counterintuitive as, in a program,  the $j$-th predicate in the body of a rule can appear in the heads of many different rules.
It is precisely to take into account this possibility that, in an execution graph, more than one node can be associated with the same rule $r$ of the program.
In this way, different nodes $v_1,\dots,v_q$ associated with the same rule $r$ can be linked with an edge labeled $j$
to different nodes ${u_1,\dots,u_q}$ whose head's predicate is the $j$-th predicate of the body of $r$.
This models that to evaluate a rule $r$ we might need to match the $j$-th predicate in the body of $r$
with facts generated by the heads of different rules.

We now define some notions on EGs that we will use throughout the paper.
For an EG $G$ for a program $P$, we denote by $\nodes{G}$ and $\edges{G}$ the sets of nodes and edges in $G$.
The depth of a node ${v \in \nodes{G}}$ is the length of the longest path that ends in $v$.  The
depth $\Depth{\EG}$ of $G$ is 0 if $G$ is the empty graph; otherwise, it is
the maximum depth of the nodes in $\nodes{V}$.


\bigchg{%
As said earlier, EGs can be used to guide the reasoning process.
In the following definition, we formalise how the reasoning over a program $P$ is carried out by following the plan encoded in an EG for $P$.
The definition assumes the following for each rule $r$ in $P$:
(i) $r$ is of the form ${\forall \vec{X} \forall \vec{Y} \bigwedge \nolimits_{i=1}^n P_i(\vec{X}_i,\vec{Y}_i) \to \exists \vec{Z} P(\vec{Y},\vec{Z})}$; and
(ii) if $r$ is intensional and is associated with a node $v$ in an EG for $P$, then the EG includes an edge of the form ${u_i \rightarrow_i v}$, for each ${1 \leq i \leq n}$.

\begin{definition}\label{definition:guided-chase}
 	Let ${(\P, \BI)}$ be a KB, $G$ be an EG for $\P$ and
 	$v$ be a node in $G$ associated with rule ${r \in P}$.
 	$v(\BI)$ includes a fact $\SH(\head{r})$, for each $h$ that is either:
 	\begin{compactitem}
 	    \item a homomorphism from the body of $r$ to $B$, if $r$ is extensional; or otherwise
 	    \item a homomorphism from the body of $r$ into ${\bigcup \nolimits_{i=1}^n u_i(B)}$
 	    so that the following holds: the restriction of $h$ over ${\vec{X}_i \cup \vec{Y}_i}$
 	    is a homomorphism from ${P_i(\vec{X}_i,\vec{Y}_i)}$ into $u_i(\BI)$, for each ${1 \leq i \leq n}$.
 	\end{compactitem}
 We pose ${\TG(B) = B \cup \bigcup_{v \in V} v(B)}$.
 \end{definition}
 }

TGs are EGs guaranteeing the correct computation of conjunctive query answering.

\begin{definition}
\label{definition:trigger-graph}
	An EG $G$ is a TG for $(P,B)$, if
	for each BCQ $Q$,
	${( \P, \BI ) \models Q}$ iff ${\TG(\BI) \models Q}$.
	$G$ is a TG for $P$, if for each base instance $B$, $G$ is a TG for $(P,B)$.
\end{definition}
TGs that depend both on $P$ and $B$ are called \emph{instance-dependent},
while TGs that depend only on $P$ are called \emph{instance-independent}.
The EGs shown in Figure~\ref{figure:trigger-graphs} are both instance-independent TGs for $P_1$.


\smallchg{%
We provide an analysis of the class of programs that admit a finite instance-independent TG denoted as \Ftg.
Theorem~\ref{theorem:relationship} summarizes the relationship between \Ftg and
the classes of programs that are bounded (\Bdd, \cite{k-boundedness}),
term-depth bounded (\Tdb, \cite{DBLP:conf/dlog/LeclereMU16}) and first-order-rewritable (\Fus, \cite{general-datalog-framework}).

\begin{restatable}{theorem}{thmrelationship}\label{theorem:relationship}
	The following hold: \P is \Ftg iff it is \Bdd; and \P is \Tdb $\cap$ \Fus iff it is \Bdd.
\end{restatable}

This result is obtained by showing that if \P is \Ftg, then it is \Bdd with
bound the maximal depth of any instance-independent TG for \P.  If it is \Bdd
with bound $k$, then the (finite) EG $G^k$, which is described after
Definition~\ref{eq:compatible-nodes}, is a TG for \P.

If a program is \Fus, then all facts that contain terms of depth at most $k$ are produced in a fixed number of chase steps.
Therefore, if it is also \Tdb, then all relevant facts in the chase are also produced in a fixed number of steps.
Finally, the undecidability of \Ftg follows from the fact that \Fus and \Ftg coincide for Datalog programs, which are always \Tdb.
See the appendix for a detailed explanation.
}

\bigchg{%
%

We conclude our analysis by showing that any KB that admits a finite model, also
admits a finite instance-dependent TG, as stated in the following statement.

\begin{restatable}{theorem}{thmTGexistence}\label{theorem:TGexistence}
        For each KB ${(P,B)}$ that admits a finite model, there exists an instance-dependent TG.
\end{restatable}

The key insight is that we can build a TG that mimics the chase. Below, we analyze the conditions
under which the same rule execution takes place both in the chase and when reasoning over a TG.
Based on this analysis we present a technique for computing instance-dependent TGs that mimic breadth-first chase variants.



Consider a rule of the form~\eqref{rule} and assume that the chase over a KB ${(P,B)}$ executes $r$ in some round $k$ by instantiating its body using the facts ${R(\vec{c}_i)}$.
Consider now a TG $G$ for ${(P,B)}$. If ${k=1}$, then this rule execution (notice that the rule has to be extensional) takes place in $G$ if there is a node $v$ associated with $r$. Otherwise, if ${k>1}$, then this rule execution takes place in $G$ if the following holds: (i) there is a node $v$ associated with $r$, (ii) each ${R(\vec{c}_i)}$ is stored in some node $u_i$ and (iii) there is an incoming edge ${u_{i} \rightarrow_i v}$,
for each ${1 \leq i \leq n}$. We refer to each combination of nodes of depth ${< k}$ whose facts
may instantiate the body of a rule $r$ when reasoning over an EG, as $k$-compatible nodes for $r$:

\begin{definition}\label{eq:compatible-nodes}
	Let $P$ be a program, $r$ be an intensional rule in $P$ and $G$ be an EG for $P$.
	A combination of $n$ (not-necessarily distinct) nodes ${(u_{1},\dots,u_{n})}$ from $G$ is $k$-compatible with
	$r$, where ${k \geq 2}$ is an integer, if:
	\begin{compactitem}
		\item the predicate in the head of $u_i$ is $R_i$;
		\item the depth of each $u_i$ is less than $k$; and
		\item at least one node in ${(u_{1},\dots,u_{n})}$ is of depth ${k-1}$.
	\end{compactitem}
\end{definition}

The above ideas are summarized in an iterative procedure, which builds at each step $k$ a graph $G^k$:
\begin{compactitem}
	\item (\textbf{Base step}) if ${k=1}$, then for each extensional rule $r$ add to $G^k$ a node $v$ associated with $r$.

	\item (\textbf{Inductive step}) otherwise, for each intensional rule $r$ and each combination of nodes ${(u_{1},\dots,u_{n})}$ from $G^{k-1}$ that is $k$-compatible with $r$,
	add to $G^k$: (i) a fresh node $v$ associated with $r$ and (ii) an edge ${u_{i} \rightarrow_i v}$, for each ${1 \leq i \leq n}$.
\end{compactitem}
The inductive step ensures that $G^k$ encodes each rule execution that takes place in the $k$-th chase round.


So far, we did not specify when the TG computation process stops. 
When $P$ is Datalog, we can stop when ${G^{k-1}(B) = G^{k}(B)}$. Otherwise, we can employ the termination criterion of the equivalent chase, e.g., ${G^{k-1}(B) \models G^{k}(B)}$, or of the restricted chase.
}

%% file: sections/computing-trigger-graphs.tex
\section{TGs for Linear Programs}\label{section:graph}

\smallchg{In the previous section, we outlined a procedure to compute
\emph{instance-dependent TGs} that mimics the chase. Now, } we propose an
algorithm {for} computing \emph{instance-independent} TGs for linear programs.

\bigchg{%
Our technique is based on two ideas.
The first one is that, for each base instance $B$, the result of chasing $B$ using a linear program $P$
is logically equivalent to the union of the instances computed when chasing each single fact in $B$ using $P$.

The second idea is based on \emph{pattern-isomorphic} facts: facts with the same predicate name and for which there is a bijection between their constants.
For example, ${R(1,2,3)}$ is pattern-isomorphic to ${R(5,6,7)}$ but not to ${R(9,9,8)}$.
We can see that two different pattern-isomorphic facts will have the same linear rules executed in the same order during chasing.
We denote by $\mathcal{H}(P)$ a set of facts formed over the extensional predicates in a program $P$, 
where no fact ${f_1 \in \mathcal{H}(P)}$ is pattern isomorphic to some other fact ${f_2 \in \mathcal{H}(P)}$.

Algorithm~\ref{alg:Trigger-Graph-Computation-Linear} combines these two ideas:
it runs the chase for each fact in $\mathcal{H}(P)$ 
then tracks the rule executions and (iii) based on these rule executions it computes a TG.
In particular, for each fact $f_2$ that is derived after executing a rule $r$ over $f_1$, Algorithm~\ref{alg:Trigger-Graph-Computation-Linear}
will create a fresh node $u$ and associate it with rule $r$,
lines~\ref{alg:Trigger-Graph-Computation-Linear:tgraph:node:start}--\ref{alg:Trigger-Graph-Computation-Linear:tgraph:node:end}.
The mapping $\mu$ associates nodes with rule executions.
Then, the algorithm adds edges between the nodes based on the sequences of rule executions that took place during chasing,
lines~\ref{alg:Trigger-Graph-Computation-Linear:tgraph:edge:start}--\ref{alg:Trigger-Graph-Computation-Linear:tgraph:edge:end}.

Algorithm~\ref{alg:Trigger-Graph-Computation-Linear} is (implicitly) parameterized by the chase variant.
The results below are based on the equivalent chase, as it ensures termination for \Fes programs.
}

\begin{restatable}{theorem}{thmtriggerGraphLinear}\label{theorem:triggerGraphLinear}
    For any linear program $P$ that is \Fes, $\tgraphLinear(P)$ is a TG for $P$.
\end{restatable}

Algorithm~\ref{alg:Trigger-Graph-Computation-Linear} has a double-exponential overhead.

\begin{restatable}{theorem}{thmtriggerGraphLinearcomplexity}\label{theorem:triggerGraphLinear-complexity}
	The execution time of Algorithm~\ref{alg:Trigger-Graph-Computation-Linear} for \Fes programs is double exponential in the input program $P$.
	If the arity of the predicates in $P$ is bounded, the execution time is (single) exponential.
\end{restatable}

\begin{algorithm}[tb]
\begin{scriptsize}
\caption{$\tgraphLinear(P)$}\label{alg:Trigger-Graph-Computation-Linear}
\begin{algorithmic}[1]
	\State \bigchg{Let ${G}$ be an empty EG}
	\For{\bigchg{\textbf{each} $f \in \mathcal{H}(P)$}}                                                                                  						\label{alg:Trigger-Graph-Computation-Linear:tgraph:start}
	      \State \bigchg{$\Gamma$ is an empty $EG$;\; $\mu$ is the empty mapping}													\label{alg:Trigger-Graph-Computation-Linear:tgraph:Gamma:init}
		\For{\bigchg{\textbf{each} ${f_1 \rightarrow_r f_2 \in \chaseGraph(P,\{f\})}$}} 												\label{alg:Trigger-Graph-Computation-Linear:tgraph:node:start}
	            	\State \bigchg{\textbf{add} a fresh node $u$ to ${\nodes{\Gamma}}$ with $\rul{u} \defeq r$}
			\State \bigchg{${\mu(u) \defeq f_1 \rightarrow_r f_2}$}
		 \EndFor																												\label{alg:Trigger-Graph-Computation-Linear:tgraph:node:end}
		\For{\bigchg{\textbf{each} $v,u \in \nodes{\Gamma}$}} 																	\label{alg:Trigger-Graph-Computation-Linear:tgraph:edge:start}
	           	\If{\bigchg{${\mu(v) = f_1 \rightarrow_r f_2}$ \textbf{and} ${\mu(u) = f_2 \rightarrow_{r'} f_3}$ }}
	                   \State \bigchg{\textbf{add} $v \rightarrow_1 u$ to $\edges{\Gamma}$}
	              \EndIf
		\EndFor																													\label{alg:Trigger-Graph-Computation-Linear:tgraph:edge:end}
		\State \bigchg{${G \defeq G \cup \Gamma}$}
	\EndFor																														\label{alg:Trigger-Graph-Computation-Linear:tgraph:end}
	\State \bigchg{\textbf{return} $G$}
\end{algorithmic}
\end{scriptsize}
\end{algorithm}

%% file: sections/linear-minimization.tex
\subsection{Minimizing TGs for linear programs} \label{section:minimization}

The TGs computed by Algorithm~\ref{alg:Trigger-Graph-Computation-Linear} may comprise nodes which can be deleted without compromising query answering. Let us return to Example~\ref{example:running:demo} and to the TG $G_1$ from Figure~\ref{figure:trigger-graphs}: we can safely ignore the facts associated with the node $u_2$ from $G_1$ and still preserve the answers to all queries over ${(P_1,B)}$. In this section, we show a technique for minimizing TGs for linear programs.

Our minimization algorithm is based on the following.
Consider a TG $G$ for a linear program $P$, a base instance $B$ of $P$ and the query
${Q(X) \leftarrow R(X,Y) \wedge S(Y,Z,Z)}$. Assume that there exists a homomorphism from the body of the query into the facts ${f_1 = R(c_1, \lnu_1)}$ and ${f_2 = S(\lnu_1,\lnu_2, \lnu_2)}$ and that ${f_1 \in v (B)}$ and ${f_2 \in u(B)}$ with $v,u$ being two nodes of $G$. Since $\lnu_1$ is shared among two different facts associated with two different nodes, it is safe to remove $u$ if there is another node ${u' \in \nodes{G}}$ whose instance $u'(B)$ includes a fact of the form ${ S(\lnu_1,\lnu'_2, \lnu'_2)}$. Equivalently, it is safe to remove $u$ if there exists a homomorphism from $u(B)$ into $u'(B)$ that maps to itself each null occurring both in $u(B)$ and $u'(B)$. Since a null can occur both in $u(B)$ and in $u'(B)$ if $u,u'$ share a common ancestor we can rephrase the previous statement as follows: we can remove $u(B)$ if there exists a homomorphism from $u(B)$ into $u'(B)$ preserving each null (from $u(B)$) that also occurs in some $w(B)$ with $w$ being an ancestor of $u$ in $G$.
We refer to such homomorphisms as \emph{preserving homomorphisms}:

\begin{definition}\label{definition:preserving}
	Let $G$ be a TG for a program $P$, ${u,v \in \nodes{G}}$ and $B$ be a base instance.
	A homomorphism from $u(B)$ into $v(B)$ is \emph{preserving}, if it maps to itself
	each null occurring in some $u'(B)$ with $u'$ being an ancestor of $u$.
\end{definition}

It suffices to consider only the \smallchg{facts} in $\mathcal{H}(P)$ to verify the existence of preserving homomorphisms.

\begin{restatable}{lemma}{lemmalinearredundant}\label{lemma:linear:redundant}
    Let $P$ be a linear program, $G$ be an EG for $P$ and ${u,v \in \nodes{G}}$.
    Then, there exists a preserving homomorphism from $u(B)$ into $v(B)$ for each base instance $B$,
    iff there exists a preserving homomorphism from $u(\{f\})$ into $v(\{f\})$,
    for each fact ${f \in \mathcal{H}(P)}$.
\end{restatable}

From Definition~\ref{definition:preserving} and from
Lemma~\ref{lemma:linear:redundant} it follows that a node $v$ of a TG can be ``ignored'' for query answering
if there exists a node $v'$ and a preserving homomorphism from $v(\{f\})$ into $v'(\{f\})$, for each ${f \in \mathcal{H}(P)}$.
If the above holds, then we say that \emph{$v$ is dominated by $v'$}.
The above implies a strategy \smallchg{to reduce the size of TGs}.

\begin{definition}\label{definition:minimization}
	For a TG $G$ for a linear program $P$, the EG $\mathsf{minLinear}(G)$ is obtained by exhaustively applying the steps:
	(i) choose a pair of nodes $v,v'$ from $G$ where $v$ is dominated by $v'$, (ii) remove $v$ from $\nodes{G}$; and (iii)
	add an edge ${v' \rightarrow_1 u}$, for each edge ${v \rightarrow_1 u}$ from $\edges{G}$.
\end{definition}




\smallchg{The minimization procedure described in Definition~\ref{definition:minimization} is correct:
given a TG for a linear program $P$, the output of $\mathsf{minLinear}$ is still a TG for $P$.}

\begin{restatable}{theorem}{thmminimizeexists}\label{theorem:minimizeexists}
    For a TG $G$ for a linear program $P$, $\mathsf{minLinear}(G)$ is a TG for $P$.
\end{restatable}


\bigchg{
We present an example demonstrating the TG computation and minimizes techniques described above.

\begin{example}\label{example:Trigger-Graph-Computation-Linear} Recall
    Example~\ref{example:running:demo}.  Since $\rs$ is the only extensional
    predicate in $P_1$, ${\mathcal{H}(P_1)}$ will include two facts, say
    ${\rs(c_1,c_2)}$ and ${\rs(c_3,c_3)}$, where $c_1$, $c_2$ and $c_3$ are constants.
    Algorithm~\ref{alg:Trigger-Graph-Computation-Linear} computes a TG by
    tracking the rule executions that take place when chasing each fact in
    ${\mathcal{H}(P_1)}$.  For example, when considering ${\rs(c_1,c_2)}$, the
    graph $\Gamma$ computed in lines
    \ref{alg:Trigger-Graph-Computation-Linear:tgraph:Gamma:init}--\ref{alg:Trigger-Graph-Computation-Linear:tgraph:edge:end}
    will be the TG $G_1$ from Figure~\ref{figure:trigger-graphs}(b), where nodes are denoted as $u_1$, $u_2$, and $u_3$.

	Let us now focus on the minimization algorithm. To minimize $G_1$, we need to identify nodes that are dominated by others.
	Recall that a node $u$ in $G_1$ is dominated by a node $v$, if for each $f$ in ${\mathcal{H}(P_1)}$,
	there exists a preserving homomorphism from ${u(\{f\})}$ into ${v(\{f\})}$.
	Based on the above, we can see that $u_2$ is dominated by $u_3$. For example, when ${B^* = \{\rs(c_1,c_2)\}}$,
	there exists a preserving homomorphism from ${u_2(B^*) = \{R(c_2,c_1,\lnu_1)\}}$ into
	${u_3(B^*) = \{R(c_2,c_1,c_1)\}}$
	mapping $\lnu_1$ to $c_1$. Since $u_2$ is dominated by $u_3$, the minimization process eliminates $u_2$ from $G_1$.
	The result is the TG $G_2$ from Figure~\ref{figure:trigger-graphs}(c), since no other node in $G_2$ is dominated.
\end{example}
}

%% file: sections/online.tex
\section{Optimizing TGs for Datalog}\label{section:online}

\bigchg{There are cases where we cannot compute instance-independent TG, e.g.,
    for Datalog programs that are not also in \Ftg class. In
    such cases, we can still create an instance-dependent TG using the procedure
    outlined in Section~\ref{section:trigger-graph}. In this section, we present
    two optimizations to this procedure which avoid redundant computations.
    These optimizations work with Datalog programs; thus also with non-linear rules.

    %
}

\subsection{Eliminating redundant nodes}\label{section:online:minimization}

Our first technique is based on a simple observation. Consider a node $v$ of a TG $G$.
Assume that $v$ is associated with the rule ${a(X,Y,Z) \rightarrow A(Y,X)}$ with $a$ being extensional.
We can see that for each base instance $B$ and each fact ${a(\sigma(X), \sigma(Y), \sigma(Z))}$ in $B$,
where $\sigma$ is a variable substitution,
the fact ${A(\sigma(Y), \sigma(X))}$ is in $v(B)$.
Equivalently, for each answer $\sigma$ to ${Q(Y,X) \leftarrow a(X,Y,Z)}$,
a
fact ${A(\sigma(Y), \sigma(X))}$ is associated with $v(B)$.
The above can be generalized.
Consider a node $v$ of a TG $G$ such that $\rul{v}$ is $\bigwedge \nolimits_{i = i}^n A_i(\vec{Y}_i) \rightarrow A(\vec{X})$.
The facts in $v(B)$ can be obtained by
(i) computing the rewriting of the query ${Q(\vec{X}) \leftarrow \bigwedge
\nolimits_{i = i}^n A_i(\vec{Y}_i)}$ w.r.t.\ the rules in \smallchg{the ancestors of $v$} up to the extensional predicates;
(ii) evaluating the rewritten query over $B$; and
(iii) adding $A(\vec{t})$ to $v(B)$, for each answer $\vec{t}$ to the rewritten query over $B$-- recall that we denote answers either as substitutions or as tuples, cf. Section~\ref{section:preliminaries}.
We refer to ${Q(\vec{X}) \leftarrow \bigwedge \nolimits_{i = i}^n A_i(\vec{Y}_i)}$ as the \emph{characteristic query} of~$v$.

This observation suggests we can use query containment tests to identify nodes that can be safely removed from TGs (and EGs).
Intuitively, the na\"ive algorithm above can be modified so that, at each step $i$,
right after computing \bigchg{$G^{i}$}, and before computing \bigchg{$G^{i}(B)$},
we eliminate each node $u$ if the EG-guided rewriting over of the characteristic query of $u$ is contained in the EG-guided rewriting of the characteristic query of another node $v$.

Below, we formalize the notion of EG-rewritings, then we show the correspondence between the answers to EG-rewritings
and the facts associated with the nodes, and we finish with an algorithm eliminating nodes from~TGs.

\begin{definition}\label{definition:EP-rewriting}
	Let $v$ be a node in an EG $G$ for a Datalog program. Let $\rul{v}$ be ${\bigwedge \nolimits_{i = 1}^n A_i \rightarrow R(\vec{Y})}$.
	The EG-rewriting of $v$, denoted as ${\rew(v)}$, is the CQ computed as
    follows (w.l.o.g.\ no pair of rules $\rul{u}$ and $\rul{v}$ with ${u, v \in \nodes{G}}$ and ${u \neq v}$ shares variables):
	\begin{compactitem}
		\item form ${Q(\vec{Y}) \leftarrow R(\vec{Y})}$;
		associate $R(\vec{Y})$ with $v$;
		\item repeat the following rewriting step until no intensional atom is left in $\body{Q}$:
            (i) choose an \emph{intensional} atom ${\alpha \in \body{Q}}$;
            (ii) compute the MGU $\theta$ of ${\{\head{{u}},\alpha\}}$, where $u$ is the node associated with $\alpha$;
            (iii) replace $\alpha$ in $\body{Q}$ with $\body{{u}}$ and apply $\theta$ on the resulting $Q$;
            (iv) associate ${\theta(B_j)}$ in $\body{Q}$ with the node $w_j$, where
            $B_j$ is the $j$-th atom in $\body{{u}}$ and ${w_j \rightarrow_j u \in \edges{G}}$.
	\end{compactitem}
\end{definition}

The rewriting algorithm described in Definition~\ref{definition:EP-rewriting} is
a variant of the rewriting algorithm in \cite{DBLP:journals/tods/GottlobOP14}.
Our difference from \cite{DBLP:journals/tods/GottlobOP14} is that at each step of the rewriting process,
we consider only the rule $\rul{u}$ with $u$ being the
node with which $\alpha$ is associated with.

There is a correspondence between the answers to the nodes' EG-rewritings with the facts stored in the nodes.

\begin{restatable}{lemma}{thmEP}\label{lemma:EP-rewriting}
	Let $G$ be an EG for a Datalog program $P$ and $B$ be a base instance of $P$.
	Then for each ${v \in \nodes{G}}$ we have: $v(B)$ includes exactly a fact ${A(\vec{t})}$ with $A$ being the head predicate of $\rul{v}$,
	for each answer $\vec{t}$ to the EG-rewriting of $v$ on $B$.
\end{restatable}

Our algorithm for removing nodes from EGs is below.

\begin{definition}\label{definition:minimizationDatalog}
	The EG $\mathsf{minDatalog}(G)$ is obtained from an EG $G$ for a program $P$ by exhaustively applying these steps:
	for each pair of nodes $u$ and $v$ such that (i) the depth of $v$ is equal or larger than that of $u$,
	(ii) the predicates of $\head{\rul{v}}$ and of $\head{\rul{u}}$ are the same and (iii) the EG-rewriting of $v$
	 is contained in the EG-rewriting of $u$:
	(a) remove the node $v$ from $\nodes{G}$, and (b) add an edge ${u \rightarrow_j w}$, for each edge ${v \rightarrow_j w}$ occurring in $G$.
\end{definition}

\smallchg{The minimization technique of Definition~\ref{definition:minimizationDatalog} can be proven sound and to produce a TG with fewest nodes.}


\begin{restatable}{theorem}{thmminimizeDatalog}\label{theorem:minimizeDatalog}
    \smallchg{If $G$ is a TG for a Datalog program $P$, then $\mathsf{minDatalog}(G)$ is a minimum size TG for $P$.}
\end{restatable}

{
Deciding whether a TG of a Datalog program is of minimum size \smallchg{can be proven} co-NP-complete.
\smallchg{The problem's hardness lies is the necessity of performing} query containment tests, carried out via homomorphism tests, which require exponential time on deterministic machines (unless $P = NP$) \cite{Chandra:1977:OIC:800105.803397}.
}%
\smallchg{%
This hardness result supports the optimality of $\mathsf{minDatalog}$ in terms of complexity.
}

\begin{restatable}{theorem}{complexityMinimumDatalog}\label{theorem:complexityMinimumDatalog}
For a Datalog program $P$ and a TG $G$ for $P$, deciding whether $G$ is a TG of minimum size for $P$ is co-NP-complete.
\end{restatable}

\subsection{A more efficient rule execution strategy}\label{section:online:rule-execution}

\begin{figure}[t]
	\centering
	\includegraphics[width=0.94\columnwidth]{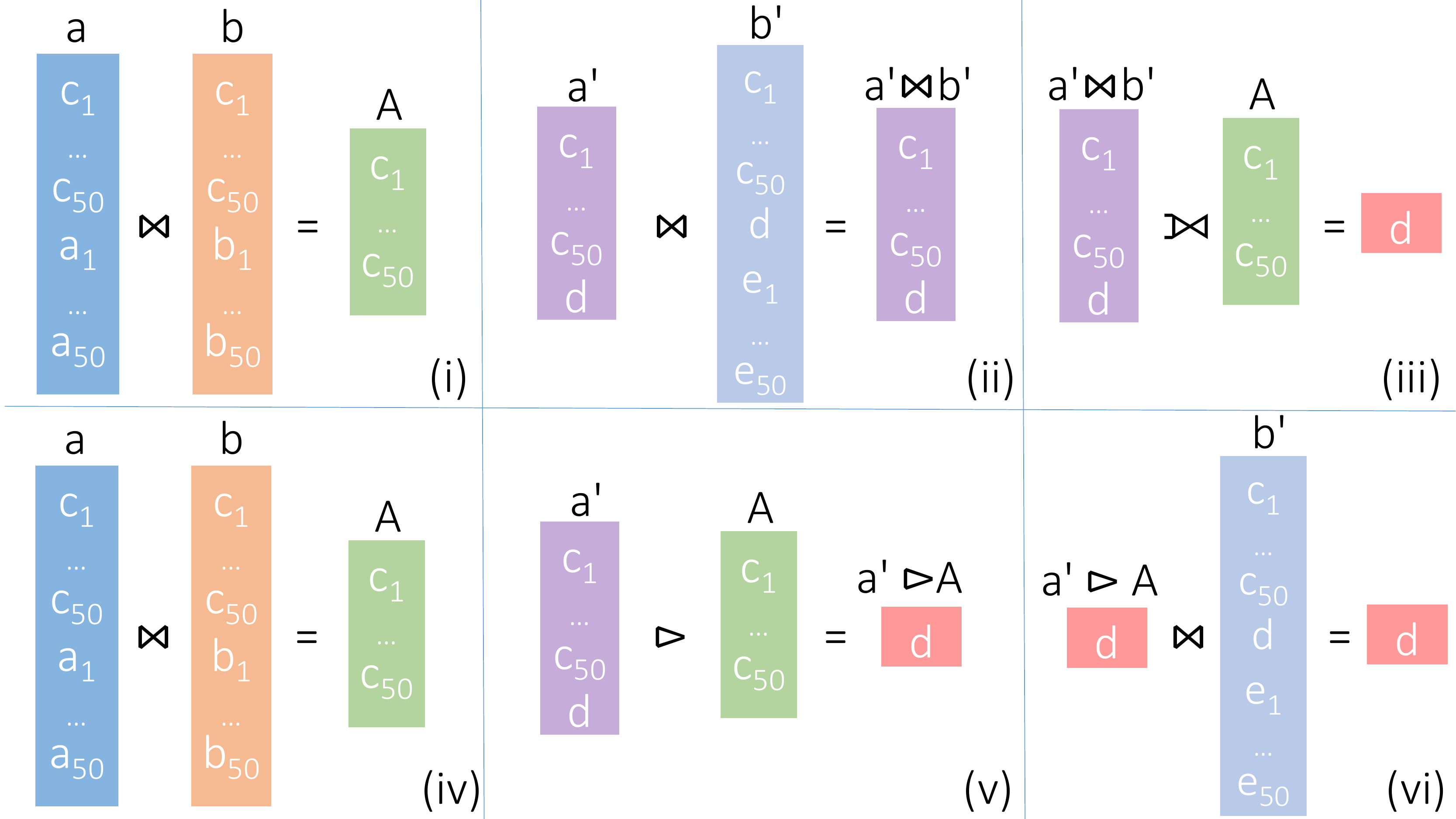} 
	\caption{Different strategies for executing the rules from $P_2$.}\label{figure:online-optimization}
\end{figure}

EG-rewritings can be further used to optimize the execution of the rules, as
shown in the example below.

\begin{example}\label{example:online-optimization}
	Consider the program $P_2$
	\begin{align*}
		a(X) \wedge b(X) &\rightarrow A(X)~~(r_8) \\
		a'(X) \wedge b'(X) &\rightarrow A(X)~~(r_9)
	\end{align*}
	where $a$, $a'$, $b$ and $b'$ are extensional predicates.
	We denote by $\mathsf{a}$, $\mathsf{a}'$, $\mathsf{b}$ and $\mathsf{b}'$
	the relations storing the tuples of the corresponding predicates in the input instance.
	The data of each relation are shown in Figure~\ref{figure:online-optimization}.

    The upper part of Figure~\ref{figure:online-optimization} shows the steps
    involved when executing $r_8$ and $r_9$ using the chase: (i) shows the
    joins involved when executing $r_8$; (ii)--(iii) show the joins involved when
    executing $r_9$: (ii) shows the join to compute $\body{r_9}$ while (iii)
    shows the outer join involved when checking whether the conclusions of $r_9$
    have been previously derived.  Assuming that the cost of executing each join
    is the cost of scanning the smallest relation, the total cost of the chase
    is: 100 (step (i)) + 51 (step (ii)) + 50 (step (iii))=201.

    The lower part of Figure~\ref{figure:online-optimization} shows a more
    efficient strategy.  The execution of $r_8$ stays
    the same (step (iv)), while for $r_9$ we first compute all
    tuples that are in $\mathsf{a}'$ but not in $\mathsf{A}$ (step (v)) and
    use ${\mathsf{a}' \setminus \mathsf{A}}$ to restrict the tuples
    instantiating the body of $r_9$ (step (vi)).  The intuition is that the
    tuples of $\mathsf{a}'$ that are already in $\mathsf{A}$ will be discarded,
    so it is not worth considering them when instantiating the body of $r_9$.
    The total cost of this strategy is: 100 (step (iv)) + 51 (step (v)) + 1
    (step (vi))=152.

\end{example}

Example~\ref{example:online-optimization} suggests a way to optimize the execution of the rules,
which reduces the cost of instantiating the rule bodies.
This is achieved by considering only the instantiations leading to the derivation of new conclusions.
Our new rule execution strategy is described below.

\begin{definition}\label{definition:EGexec}
Let $v$ be a node of an EG $G$ for a Datalog program $P$, $B$ be a base instance and ${I \subseteq G(B)}$.
Let ${A(\vec{X})}$ be the head atom of $\rul{v}$ and let
${Q(\vec{Y}) \leftarrow \bigwedge \nolimits_{i = 1}^n f_i}$ be the EG-rewriting of $v$.
\noindent
	The computation of $v(B)$ \emph{under $I$}, denoted as $v(B,I)$, is:
	\begin{compactenum}
		\item pick $m \geq 1$ atoms ${f_{i_1}, \dots, f_{i_m}}$ from the body of $Q$
		whose variables include all variables in ${\vec{Y}}$ and form ${Q'(\vec{Y}) \leftarrow f_{i_1} \wedge \dots \wedge f_{i_m}}$;

		\item compute $v(B)$ as in Definition~\ref{definition:guided-chase}, however
		restrict to homomorphisms $h$ for which (i) ${h(\vec{X})}$ is an answer to $Q'$ on $B$ and (ii) ${A(h(\vec{X})) \not \in I}$.
	\end{compactenum}
\end{definition}

To help us understand Definition~\ref{definition:EGexec},
let us apply it to
Example~\ref{example:online-optimization}. We have ${Q'(X) \leftarrow a'(X)}$.
The antijoin between $Q'$ and $A$ (step (v) of Figure~\ref{figure:online-optimization})
corresponds to restricting to homomorphisms that are answers to $Q'$ (step (2.i) of Definition~\ref{definition:EGexec}),
but are not in $I$ (step (2.ii) of  Definition~\ref{definition:EGexec}).
In our implementation, we pick one extensional atom ($m=1$) in step (1).
To pick this atom, we consider each $f_i$ in the body of
$\rew(v)$, then compute the join as in step (v) of Example~\ref{example:online-optimization}
between a subset of the $f_i$-tuples and the $A$-tuples in $I$
and finally, choose the $f_i$ leading to the highest join output.

\begin{algorithm}[tb]
\caption{$\mathsf{TGmat}(P, B)$}\label{alg:online}
\begin{scriptsize}
\begin{algorithmic}[1]
	\State ${k \defeq 0}$; \; $G^0$ is the empty graph; \; $I^0 \defeq \emptyset$
	\Do 																																						\label{algorithm:while:begin}
		\State ${k \defeq k + 1}$;  \; $I^k \defeq I^{k-1}$
		\State \bigchg{Compute $G^k$ starting from $G^{k-1}$ as in Section~\ref{section:trigger-graph}}
		\State $G^k \defeq \mathsf{minDatalog}(G^k)$ 																											\label{algorithm:prune}
		\For{\textbf{each} node $v$ of depth $k$}
			\State \textbf{add} $v(B,I^{k-1})$ (cf. Definition~\ref{definition:EGexec}) to $I^{k}$																	\label{algorithm:optExec}
		\EndFor
	\doWhile{$I^{k} \neq I^{k-1}$}																		     														\label{algorithm:while:end}
	\State \textbf{return} $I^{\infty}$
\end{algorithmic}
\end{scriptsize}
\end{algorithm}

We summarize TG-guided reasoning for
Datalog programs in Algorithm~\ref{alg:online}. Correctness is stated below.

\begin{restatable}{theorem}{thmTGmat}\label{theorem:TGmat}
	For a Datalog program $P$ and a base instance $B$, $\mathsf{TGmat}(P, B) = \Chase{P,B}$.
\end{restatable}

%% file: sections/evaluation.tex

\section{Evaluation} \label{section:evaluation}

We implemented Algorithm~\ref{alg:Trigger-Graph-Computation-Linear}, TG-guided
reasoning over a fixed TG (Def.~\ref{definition:guided-chase}) and
Algorithm~\ref{alg:online} in a new open-source reasoner called \emph{\sys{}}. \sys{} is a
fork of VLog~\cite{vlog_ijcar} that shares the same code for handling the
extensional relations while the code for reasoning
is entirely novel.

We consider \smallchg{three} performance measures: the absolute reasoning runtime,
\smallchg{the peak RAM consumption observed at reasoning time}, and the total
number of triggers. The last measure is added because it reflects the ability of
TGs to reduce the number of redundant rule executions and it is robust to most
implementation choices.


\subsection{Testbed}

\leanparagraph{Systems}
\smallchg{
We compared against the following systems:
\begin{compactitem}
	\item VLog, as, to our knowledge, is the most efficient system both time- and memory- wise \cite{vlog,vlog_ijcar};
	\item the latest public release of RDFox from \cite{RDFox-src} as it outperforms
	all chase engines tested against \chaseBench \cite{benchmarking-chase}: ChaseFun, DEMo~\cite{Pichler2009}, 
	LLunatic \cite{Geerts2014b}, PDQ \cite{usvldb14} and Pegasus \cite{meier};
	\item the commercial state of the art chase engine COM (name is anonymized due to licensing restrictions);
	\item Inferray, an RDFS reasoner that uses a columnar layout and that
        outperforms RDFox \cite{Inferray}; and
	\item WebPIE, another high-performance RDFS reasoner that runs over Hadoop
        \cite{webpie}.
\end{compactitem}

We ran VLog, RDFox and the commercial chase engine COM using their most efficient chase implementations.
For VLog, this is the restricted chase, while for RDFox and COM this is the Skolem one \cite{benchmarking-chase}.
All engines ran using a single thread.
We could not obtain access to the Vadalog~\cite{vadalog} binaries.
However, we perform an indirect comparison against Vadalog: we both compare
against RDFox using the ChaseBench scenarios from \cite{benchmarking-chase}.

}


\input{sections/table-benchmarks}
\input{sections/table-linear-datalog}

\leanparagraph{Benchmarks} To asses the performance of \sys{} on linear and
Datalog scenarios, we considered benchmarks previously used to evaluate the
performance of reasoning engines including VLog and RDFox: {LUBM} \cite{lubm}
and {UOBM}~\cite{uobm} are synthetic benchmarks; {DBpedia}~\cite{dbpedia}
\smallchg{(v2014, available
online\footnote{\url{https://www.cs.ox.ac.uk/isg/tools/RDFox/2014/AAAI/input/DBpedia/ttl/}})}
is a KG extracted from Wikipedia; {Claros}~\cite{claros} and
{Reactome}~\cite{reactome} are real-world
ontologies\smallchg{\footnote{\smallchg{Both
datasets are available in our code repository.}}}. With both VLog and
\sys{}, the KBs are stored with the RDF engine Trident~\cite{trident}.

\leanparagraph{Linear scenarios} \smallchg{Linear scenarios were created using
LUBM, UOBM, DBpedia, Claros and Reactome.} For the first four KBs, we considered
the linear rules returned by translating the OWL ontologies in each KB using the
method described by~\cite{zhou_making_2013}, which was the technique used for
evaluating our competitors~\cite{motik_parallel_2014,vlog}. \smallchg{This method converts
an OWL ontology $\cO$ into a Datalog program $P_L$ such that $\cO \models P_L$.
For instance, the OWL axiom $A\sqsubseteq B$ (concept inclusion) can be
translated into the rule $A(X) \to B(X)$. This technique is ideal for our
purposes since this subset is what is mostly supported by RDF
reasoners~\cite{motik_parallel_2014}. Here, the subscript ``L'' stands for
``lower bound''. In fact, not every ontology can be fully captured by Datalog
(e.g., ontologies that are not in OWL 2 RL) and in such cases the
translation captures a subset of all possible derivations.}

For Reactome, we considered the subset of linear rules from the program used
in~\cite{vlog_ijcar}. The programs for the first four KBs do not include any
existential rules while the program for Reactome does. Linear scenarios are
suffixed by ``LI'', e.g., LUBM-LI.

\leanparagraph{Datalog scenarios} \smallchg{Datalog scenarios were created using
LUBM, UOBM, DBpedia and Claros, as Reactome includes non-Datalog rules only}.
LUBM comes with a generator, which allows controlling the size of the base
instance by fixing the number of different universities $X$ in the instance.
One university roughly corresponds to 132k facts. In our experiments, we set $X$
to the following values: 125, 1k, 2k, 4k, 8k, 32k, 64k, 128k. This means that
our largest KB contains about 17B facts. \smallchg{As programs, we used the
entire Datalog programs (linear and non-linear) obtained
with~\cite{zhou_making_2013} as described above}. These programs are suffixed by
``L''. For Claros and LUBM, we used two additional programs, suffixed by ``LE'',
\smallchg{created by~\cite{motik_parallel_2014} as harder benchmarks}. These
programs extend the ``L'' ones with extra rules, such as the transitive and
symmetric rules for \emph{owl:sameAs}. The relationship between the various
rulesets is ${LI \subset L \subset LE}$.


\smallchg{
\leanparagraph{ChaseBench scenarios} \chaseBench{} was introduced for evaluating
the performance of chase engines~\cite{benchmarking-chase}. The benchmark comes
with four different families of scenarios. Out of these four families, we
focused on the iBench scenarios, namely STB-128 and ONT-256~\cite{ibench}
because they come with non-linear rules with existentials that involve many
joins and that are highly recursive.  Moreover, as we do compare against RDFox
which was the top-performing chase engine in~\cite{benchmarking-chase}, we can
use these two scenarios to indirectly compare against all the engines considered
in \cite{benchmarking-chase}.

\leanparagraph{RDFS scenarios} In the Semantic Web, it has been shown that a
large part of the inference that is possible under the RDF Schema
(RDFS)~\cite{rdfs} can be captured into a set of Datalog rules. A number of
works have focused on the execution of such rules. In particular, WebPIE and
more recently Inferray returned state-of-the-art performance for $\rho DF$ -- a subset of RDFS that captures its essential semantics. It is interesting
to compare the performance of \sys{}, which is a generic engine not optimized
for RDFS rules, against such ad-hoc systems. To this end, we considered YAGO~\cite{yago2} and a LUBM KB with 16.7M triples. As rules for \sys{}, we
translated the ontologies under the $\rho DF$ semantics.}

Table~\ref{tab:benchmarks} shows, for each scenario, the corresponding number of
rules and \EP-facts as well as the number of \IP-facts in the model of the KB.
With LUBM \smallchg{and the linear and Datalog scenarios}, the number of
\IP-facts is proportional to the input size, thus it is stated as \%. For
instance, with the ``LI'' rules, the output is 116\%, which means that if the
input contains 1M facts, then reasoning returns 1.16M new facts.

\smallchg{ \leanparagraph{Hardware} All experiments except the ones on
    scalability (Section~\ref{sec:scalability}) ran on an Ubuntu 16.04 Linux PC
    with Intel i7 64-bit CPU and 94.1 GiB RAM. For our experiments on
    scalability, we used a second machine with an Intel Xeon E5 and 256 GiB of
RAM due to the large sizes of the KBs.} The cost of both machines is
$<$\$5k, thus we arguably label them as commodity hardware.

\input{sections/table-extraexperiments}

\subsection{Results for linear scenarios}

Table~\ref{tab:linear} summarizes the results of our empirical evaluation for
the linear scenarios. Recall that when a program is linear and \Fes it admits a
finite TG which can be computed prior to reasoning using $\tgraphLinear$
(Algorithm~\ref{alg:Trigger-Graph-Computation-Linear}) and minimized using
$\mathsf{minLinear}$ from Definition~\ref{definition:minimization}. \smallchg{%
Columns two to seven show the runtime and the peak memory consumption for VLog,
RDFox and the commercial engine COM. The remaining columns show results related
to TG-guided reasoning. Column \emph{Comp} shows the time to compute and
minimize a TG using $\tgraphLinear$ and $\mathsf{minLinear}$. Column
\emph{Reason} shows the time to reason over the computed TG given a base
instance (i.e., apply Definition~\ref{definition:guided-chase}). Column
\emph{w/o cleaning} shows the {total runtime} if we do not filter out redundant
facts at reasoning time, while column \emph{w/ cleaning} shows the total runtime
if we additionally filter out redundancies at the end and collectively for all
the rules. Notice that in both cases the total runtime includes the time to
compute and reason over the TG (columns \emph{Comp} and \emph{Reason}).  Column
\emph{Mem} shows the peak memory consumption. As we will explain later, in
the case of linear rules, the memory consumption in \sys{} is the same both with
and without filtering out redundant facts. Finally, the last three columns \#N,
{\#E}, and D show the number of nodes, edges, and the depth (i.e., length
of the longest shortest path) in the resulting TGs.

We summarize two main conclusions
of our analysis.}

\conclude{TGs outperform the chase in terms of runtime and memory}
\smallchg{The runtime improvements over the chase vary from multiple orders of
magnitude (w/o filtering of redundancies) to almost two times (w/o filtering).
When redundancies are discarded, the vast improvements are attributed to
\emph{structure sharing}, a technique which is also implemented in VLog.

Structure sharing is about reusing the same columns to store the data of
different facts. For example, consider the rule ${R(X,Y) \rightarrow S(Y,X)}$.
Instead of creating different $S$- and $R$-facts, we can simply add a pointer
from the first column of $R$ to the second column of $S$ and a pointer from the
second column of $R$ to the first column of $S$. When a rule is linear, both
VLog and \sys{} perform structure sharing and, hence, do not allocate extra memory
to store the derived facts. Apart from the obvious benefits memory-wise,
structure sharing also provides benefits in runtime as it allows deriving new
facts without actually executing rules.  The above, along with the fact that the
facts (redundant or not) are not explicitly materialized in memory makes \sys{}
very efficient time-wise.

When redundancies are filtered out, \sys{} still outperforms the other engines: it
is multiple orders of magnitude faster than RDFox and COM and almost two times
faster than VLog (Reactome-LI). The performance improvements are attributed to a
more efficient strategy for filtering out redundancies: TGs allow filtering out
redundancies after reasoning has terminated, in contrast to the chase, which is
forced to filter out redundancies right after the derivation of new facts. This
strategy is more efficient because we can use a single n-way join rather than
multiple binary joins to remove redundancies.

With regards to memory, \sys{} has similar memory requirements with VLog,
while it is much more memory efficient than RDFox and the commercial engine COM.
}

\conclude{The TG computation overhead is small} \smallchg{
The time to compute and minimize a TG in advance of reasoning is only a small
fraction of the total runtime, see Table~\ref{tab:linear}.  We argue that even
if this time was not negligible, TG-guided reasoning would still be beneficial:
first, once a TG is computed reasoning over it is multiple times faster than the
chase and, second, the same TG can be used to reason over the same rules
independently of any changes in the database.}

\subsection{Results for Datalog and ChaseBench}
Table~\ref{tab:datalog} summarizes our results on generic (linear and
non-linear) Datalog rules. \smallchg{The last nine columns show results for
    $\mathsf{TGmat}$ (Algorithm~\ref{alg:online}). To assess the impact of
    $\mathsf{minDatalog}$ and $\mathsf{ruleExec}$, the rule execution strategy
    from Definition~\ref{definition:EGexec}, we ran $\mathsf{TGmat}$ as follows:
    without $\mathsf{minDatalog}$ or $\mathsf{ruleExec}$, column \emph{No opt};
    with $\mathsf{minDatalog}$, but without $\mathsf{ruleExec}$, column
\emph{m}; with both $\mathsf{minDatalog}$ and $\mathsf{ruleExec}$, column
\emph{m+r}. The total runtime in the last two cases includes the runtime
overhead of $\mathsf{minDatalog}$ and $\mathsf{ruleExec}$.} The last three
columns report the number of nodes, edges, and depth of the computed TGs when
both $\mathsf{minDatalog}$ or $\mathsf{ruleExec}$ are employed.
%
\smallchg{Table~\ref{tab:chaseBench} shows results for \chaseBench} while
Table~\ref{tab:triggers} shows the number of triggers for the Datalog scenarios
for VLog and \sys{} (we could not extract this information for RDFox and COM).

We summarize the main conclusions of our analysis.

\conclude{TGs outperform the chase in terms of runtime and memory} \smallchg{%
Even without any optimizations, \sys{} is faster than VLog, RDFox and COM in all
but one case. With regards to VLog, \sys{} is up to nine times faster in the
Datalog scenarios (LUBM-LE) and up to two times faster in \chaseBench{}
(ONT-256). With regards to RDFox, \sys{} is up to 20 times faster in the Datalog
scenarios (Claros-L) and up to 67 times faster in \chaseBench (ONT-256). When
all optimizations are enabled \sys{} outperforms the competitors in all cases.

We have observed that the bulk of the computation lies in the execution of the joins
involved when executing few expensive rules.
In GLog, joins are executed more efficiently than in the other engines (\sys{} uses only merge joins),
since the considered instances are smaller --recall that in TGs, the execution of
a rule associated with a node $v$ considers only the instances of the parents of $v$.
Due to the above, the optimizations do not decrease the runtime considerably.
The only exception is LUBM-L, where the optimizations half the runtime.

Continuing with the optimizations, their runtime overhead is very low:
it is 9\% of the total runtime (LUBM-L), while the overhead
of $\mathsf{minDatalog}$ is less than 1\% of the total runtime (detailed results are in the appendix).
We consider this overhead to be acceptable, since, as we shall see later,
the optimizations considerably decrease the number of triggers, a performance measure which is
robust to hardware and most implementation choices.
}

\smallchg{It is important to mention that} GLog implements the technique in
\cite{motik1} for executing transitive and symmetric rules.  \smallchg{The
    improvements brought by this technique are most visible with LUBM-LE where
    the runtime increases from 18s with this technique to 71s without it. Other
    improvements occur with UOBM-L and DBpedia-L (69\% and 57\% resp.). In any
case, even without this technique, \sys{} remains faster than its competitors in
all cases.}

\smallchg{Last, the \chaseBench experiments allow us to compare against Vadalog.
According to \cite{vadalog}, Vadalog is three times faster than RDFox on STB-128
and ONT-256.  Our empirical results show that \sys{} brings more substantial
runtime improvements: \sys{} is from 49 times to more than 67 times faster than
RDFox in those scenarios.

With regards to memory, the memory footprint of \sys{}
again is comparable to that of VLog and it is lower than that of RDFox and of COM.
}

\conclude{TGs outperform the chase in terms of the number of triggers}
\smallchg{
Table~\ref{tab:triggers} shows that the total number of triggers and, hence, the
amount of redundant computations, is considerably lower than the total number of
triggers in VLog even when the optimizations are disabled. This is due to the
different approaches employed to filter out redundancies: VLog filters out
redundancies right after the execution of each rule~\cite{vlog}, while \sys{}
performs this filtering after each round. When the optimizations are enabled,
the number of triggers further decreases: in the best case (DBpedia-L), \sys{}
computes 1.69 times fewer triggers (79M/47M).
}


\subsection{Results for RDFS scenarios} \smallchg{Table~\ref{tab:rdfs}
    summarizes the results of the RDFS scenarios where \sys{} is configured with
    both optimizations enabled. We can see that \sys{} is faster than both RDFS
    engines. With regards to Inferray, \sys{} is two orders of magnitude faster
    on LUBM and more than four times faster on YAGO. With regards to WebPIE,
    \sys{} is three orders of magnitude faster on LUBM and more than 32 times
    faster on YAGO. With regards to memory, \sys{} is more memory efficient in
all but one cases.}

\subsection{Results on scalability}
\label{sec:scalability}

\smallchg{
We used the LUBM benchmark to create several KBs with 133M, 267M, 534M, 1B, 2B,
4B, 8B, and 17B facts respectively. Table~\ref{tab:scalability} summarizes the
performance with the Datalog program LUBM-L. Columns are labeled with the size
of the input database. Each column shows the runtime, 
the peak RAM memory consumption, and the number of derived facts for each input
database. We can see that \sys{} can reason with up to 17B facts in less than 40
minutes without resorting to expensive hardware. We are not aware of any other
centralized reasoning engine that can scale up to such an extent.}



%% file: sections/table-benchmarks.tex
\begin{table}[t]
    \centering
    \begin{adjustbox}{max width=\columnwidth}
            \begin{tabular}{ p{1cm} c || ccc || ccc}
                                & & \multicolumn{3}{c}{\#Rules} & \multicolumn{3}{c}{\#\IP's}\\
                 Scenario        & \#\EP's & LI & L & LE & LI & L & LE \\
                  \hline
                \multicolumn{8}{l}{\bf Linear and Datalog scenarios} \\
                LUBM & {var.} & 163 & 170 & 182 & 116\% & 120\% & 232\% \\
                UOBM & 2.1 & 337 & 561 & NA & 3.5 & 3.9 & NA \\
                DBpedia & 29 & 4204 & 9396 & NA & 31.9 & 33.1 & NA \\
                Claros & 13.8 & 1749 & 2689 & 2749 & 65.8 & 8.9 & 548 \\
                React. & 5.6 & 259 & NA & NA & 11.3 & NA & NA \\
                \multicolumn{8}{l}{\bf \smallchg{ChaseBench scenarios}} \\
                S-128 & \smallchg{0.15} & \multicolumn{3}{c||}{\smallchg{167}} & \multicolumn{3}{c}{\smallchg{1.9}} \\
                O-256 & \smallchg{1} & \multicolumn{3}{c||}{\smallchg{529}} & \multicolumn{3}{c}{\smallchg{5.6}} \\
                \multicolumn{8}{l}{\bf \smallchg{RDFS ($\rho$DF) scenarios}} \\
                LUBM & \smallchg{16.7} & \multicolumn{3}{c||}{\smallchg{160}} & \multicolumn{3}{c}{\smallchg{18}} \\
                YAGO & \smallchg{18.2} & \multicolumn{3}{c||}{\smallchg{498016}} & \multicolumn{3}{c}{\smallchg{27}} \\
            \end{tabular}
    \end{adjustbox}
    \caption{The considered benchmarks. \#\EP's and \#\IP's absolute numbers are
    stated in millions of facts.}
    \label{tab:benchmarks}
\end{table}

%% file: sections/table-linear-datalog.tex
\begin{table*}[tb]
    \centering
    \scriptsize
    \begin{adjustbox}{max width=\textwidth}
    \smallchg{
    \begin{tabular}{ p{1.5cm}
        S[table-format=1.1,table-column-width=0.5cm]
        S[table-format=4,table-column-width=0.5cm]
        || S[table-format=2.1,table-column-width=0.6cm]
           S[table-format=4,table-column-width=0.5cm]
        || S[table-format=2.1,table-column-width=0.6cm]
           S[table-format=4,table-column-width=0.5cm]
        || S[table-format=1.3,table-column-width=0.8cm]
        S[table-format=1.3,table-column-width=0.8cm]
        S[table-format=1.3,table-column-width=1.3cm]
        S[table-format=1.1,table-column-width=1.3cm]
        S[table-format=4,table-column-width=0.5cm]
        || S[table-format=5,table-column-width=0.5cm]
        S[table-format=4,table-column-width=0.5cm]
        S[table-format=2,table-column-width=0.2cm]}
        			  & \multicolumn{2}{c}{\textbf{VLog}}
        			  & \multicolumn{2}{c}{\textbf{RDFox}}
                            & \multicolumn{2}{c}{\textbf{COM}}
                            & \multicolumn{5}{c}{\textbf{GLog}}
                            & \multicolumn{3}{c}{\textbf{TG Sizes}} \\

            \textbf{Scenario}  & {Run.} & {Mem} & {Run.} & {Mem} &
            {Run.} & {Mem} & {Comp} & {Reason} & {w/o cleaning}  & {w/
            cleaning} & {Mem} & \#N & {\#E} & {D} \\
        \hline
        LUBM-LI & 1.3 & 	1617	& 22 & 	2353	& 18.4 & 5122 		& 0.007
                & 0.2     & 0.207 & 1.1 & 1674	& 155 & 101 & 6 \\
        UOBM-LI & 0.3 & 	221		& 3.9 & 	726		& 3.3 & 3570 		&
        0.01   & 0.015 & 0.025 & 0.2 &   219 & 313 & 206 & 9\\
        DBpedia-LI & 6.9 & 	2579	& 44.1 & 3197 	& 36.3 & 3767 		& 0.448
                   & 0.776 & 1.224 & 4.5 & 2647	& 12660 & 8970 & 17\\
        Claros-LI & 5.6 & 	2870	& 78.4 & 3918	& 72.3 & 5122 		& 0.006
                  & 0.407 & 0.413 & 4.8 & 2586	& 792 & 621 & 23 \\
        React.-LI & 1.8 & 	1312	& 12.7 & 1448	& 9.9 & 4479 		& 0.002
                  & 0.329 & 0.329 & 0.9 &   1312 & 386 & 263 & 8\\
    \end{tabular} }
    \end{adjustbox}
        \caption{\smallchg{Linear scenarios. Time is in sec and memory in MB.}}
    \label{tab:linear}

    \centering
    \scriptsize
    \begin{adjustbox}{max width=\textwidth}
    \smallchg{
    \begin{tabular}{ p{1.5cm}
            S[table-format=4.1,table-column-width=0.8cm]
            S[table-format=4,table-column-width=0.5cm]
        || S[table-format=4.1,table-column-width=0.8cm]
            S[table-format=4,table-column-width=0.5cm]
        || S[table-format=4.1,table-column-width=0.8cm]
            S[table-format=4,table-column-width=0.5cm]
        || S[table-format=4.1,table-column-width=0.8cm]
            S[table-format=4.1,table-column-width=0.8cm]
            S[table-format=4.1,table-column-width=0.8cm]
        || S[table-format=4,table-column-width=0.8cm]
           S[table-format=4,table-column-width=0.8cm]
           S[table-format=4,table-column-width=0.8cm]
	|| S[table-format=4,table-column-width=0.5cm]
       S[table-format=4,table-column-width=0.5cm]
       S[table-format=2,table-column-width=0.2cm]}
        & \multicolumn{2}{c}{\textbf{VLog}}
        & \multicolumn{2}{c}{\textbf{RDFox}}
        & \multicolumn{2}{c}{\textbf{COM}}
        & \multicolumn{3}{c}{\textbf{GLog Runtime}}
        & \multicolumn{3}{c}{\textbf{GLog Memory}}
        & \multicolumn{3}{c}{\textbf{TG Sizes}} \\
        \textbf{Scenario}  & {Run.} & {Mem} & {Run.} & {Mem} &
        {Run.} & {Mem} & {No opt} & {m} & {m+r} & {No opt} & {m} & {m+r} & \#N &
        {\#E} & {D} \\
        \hline
        LUBM-L & 1.5 & 324 			& 23 & 2301 		& 20.4      & 4479
               &2.4 & 2.2 & 1.0                     & 446 & 424 & 264 & 56 & 33
               & 4\\
        LUBM-LE & 170.5 & 2725 		& 116.6 & 3140 		& 115.9    & 3610
                & 17.3 & 17.2 & 16.1               & 1340 & 1310 & 1338 & 63 &
        43  & 5\\
        UOBM-L & 7.3& 1021 			& 10 & 784 			& 10        & 4215
               & 2.6 & 2.4 & 2.6                    & 335 & 335 & 342 & 527 &
        859 & 6\\
        DBpedia-L & 41.6 & 827 			& 64.4 & 3290 		& 198.4    & 3878
                  & 20 & 19 & 19                       & 1341 & 1352 & 1339 &
        4144 & 3062 & 8\\
        Claros-L & 431 & 3170 			& 2512 & 5491 		& 2373.0  & 6453
                 & 122 & 118.3 & 119               & 6076 & 6077 & 6078 & 438 &
        404 & 9\\
        Claros-LE & 2771.8 & 11895 		& * & * 				& *          & *
                  & 1040.8 & 1012.2 & 1053.9    & 48464 & 48474 & 48455 & 1461 &
        3288 & 9\\
    \end{tabular} }
    \end{adjustbox}
    \caption{\smallchg{Datalog scenarios. Time is in sec and memory in MB. $*$
    denotes timeout after 1h.}}
    \label{tab:datalog}
\end{table*}

%% file: sections/table-extraexperiments.tex

\begin{table*}[tb]
    \scriptsize
        \raisebox{0.88em}{
        \begin{minipage}{0.62\textwidth}
        \smallchg{
            \input sections/table-chasebench.tex
        }
        \caption{\smallchg{\chaseBench scenarios (S=STB-128,O=ONT-256). Runtime
        in sec, memory in MB.}}
        \label{tab:chaseBench}
        \end{minipage}}
        \begin{minipage}{0.38\textwidth}
        \smallchg{
            \center
            \input sections/table-triggers.tex
            \caption{\smallchg{\#Triggers (millions), Datalog scenarios.}}
            \label{tab:triggers}
            \vspace{-1em}
        }
        \end{minipage}

        \raisebox{0.5em}{
        \begin{minipage}{0.5\textwidth}
        \begin{center}
        \smallchg{
        \begin{adjustbox}{max width=\textwidth}
            \input sections/table-rdfs.tex
        \end{adjustbox}
        \caption{\smallchg{RDFS scenarios (L=LUBM,Y=YAGO). Runtime in sec, memory in MB.}}
        \label{tab:rdfs}}
        \end{center}
        \end{minipage}}
        \begin{minipage}{0.5\textwidth}
            \scriptsize
            \smallchg{
            \input sections/table-scalability.tex
            \caption{\smallchg{Scalability results. Runtime in sec, memory in
            GB.}}
            \label{tab:scalability}
            }
        \end{minipage}

\end{table*}

%% file: sections/table-chasebench.tex
\begin{tabular}{ p{0.2cm}
    ||  S[table-format=1.1,table-column-width=0.5cm]
        S[table-format=4,table-column-width=0.5cm]
    || S[table-format=2.1,table-column-width=0.5cm]
        S[table-format=4,table-column-width=0.5cm]
    || S[table-format=2,table-column-width=0.5cm]
        S[table-format=4,table-column-width=0.5cm]
    || S[table-format=2.1,table-column-width=0.5cm]
        S[table-format=4,table-column-width=0.5cm]
 || S[table-format=3,table-column-width=0.4cm]
      S[table-format=2,table-column-width=0.4cm]
      S[table-format=1,table-column-width=0.4cm]
  }
    & \multicolumn{2}{S[table-format=1.3,table-column-width=1.5cm]}{\textbf{VLog}}
    & \multicolumn{2}{S[table-format=1.3,table-column-width=1.5cm]}{\textbf{RDFox}}
    & \multicolumn{2}{S[table-format=1.3,table-column-width=1.5cm]}{\textbf{COM}}
    & \multicolumn{2}{S[table-format=6.2,table-column-width=1.5cm]}{\textbf{GLog}}
    & \multicolumn{3}{S[table-format=6.2,table-column-width=1.5cm]}{\textbf{TG Sizes}}\\

    \textbf{S}  & {Run.} & {Mem} & {Run.} & {Mem} & {Run.} & {Mem} & {Run.} &
    {Mem} & \#N & {\#E} & {D} \\

    \hline
        S& 0.5  & 1350 			& 13.4 & 1747 		& 10      & 5217
         & 0.2 & 1266  & 192 & 0 & 0 \\
        O& 2.3 & 4930 			& 49 & 3997 		& 35    	& 6340
         & 1 & 4930 & 577 & 65 & 3 \\
\end{tabular}

%% file: sections/table-triggers.tex
\begin{tabular}{ p{1.4cm}
     S[table-format=2,table-column-width=0.5cm]
    S[table-format=2,table-column-width=0.5cm]
    S[table-format=2,table-column-width=0.5cm]
    S[table-format=2,table-column-width=0.5cm]}
    \textbf{Scenario}  & {\textbf{VLog}} & \multicolumn{3}{c}{\textbf{GLog}} \\
                       & & {no opt} & {m} & {m+r}\\
    \hline
    LUBM-L & 38 & 32 & 29 & 25 \\
    LUBM-LE & 239 & 100 & 98 & 93 \\
    UOBM-L & 47 & 9 & 8 & 8 \\
    DBpedia-L & 79 & 63 & 61 & 47 \\
    Claros-L & 286 & 218 & 195 & 185 \\
    Claros-LE & 1099 & 1072 & 1049 & 1039 \\
\end{tabular}

%% file: sections/table-rdfs.tex
%
%

\begin{tabular}{ p{0.2cm}
        || S[table-format=3,table-column-width=0.5cm]
            S[table-format=4,table-column-width=0.5cm]
        || S[table-format=3.1,table-column-width=0.5cm]
            S[table-format=5,table-column-width=0.6cm]
        || S[table-format=2.1,table-column-width=0.5cm]
           S[table-format=4,table-column-width=0.5cm]
        || S[table-format=4,table-column-width=0.5cm]
           S[table-format=4,table-column-width=0.5cm]
           S[table-format=1,table-column-width=0.3cm]
	  }
        & \multicolumn{2}{S[table-format=6.2,table-column-width=1cm]}{\textbf{WebPIE}}
        &
        \multicolumn{2}{S[table-format=6.2,table-column-width=1.5cm]}{\textbf{Inferray}}
        & \multicolumn{2}{c}{\textbf{GLog}}
        & \multicolumn{3}{S[table-format=6.2,table-column-width=1.5cm]}{\textbf{TG Sizes}}\\

        \textbf{S}  & {Run.} & {Mem} & {Run.} & {Mem} & {Run.} & {Mem} & \#N &
        {\#E} & {D} \\

        \hline
        L 
              & 338  & 1124  			& 39 & 7000 			
              & 0.3  		
              & 186 & 53 & 25 & 4 \\
        Y 
           & 745  & 1075 			& 116.6 & 14000 		
              & 25 
              &	1603 & {1.07M}
              & {888k} & 20
              \\
\end{tabular}

%% file: sections/table-scalability.tex
%
%

\begin{tabular}{ p{0.6cm} c c c c c c c c }
    & {133M}  & {267M} & {534M} & {1B} & {2B} & {4B} & {8B} & {17B} \\
    \hline
    Run. & 13 & 27 & 56 & 203 & 226 & 520 & 993 & 2272 \\
    Mem & 1 & 3 & 6 & 23 & 34 & 49 & 98 & 174 \\
    \#\IP's & 160M & 320M & 641M & 1B & 2B & 5B & 10B & 20B\\
\end{tabular}

%% file: sections/related.tex
\section{Related work}\label{section:related}

\smallchg{

One approach to improve the reasoning performance is to parallelize the
execution of the rules. RDFox proposes a parallelization technique for Datalog
materialization with mostly lock-free data insertion. Parallelization has been
also been studied for reasoning over RDFS and OWL ontologies. For example,
WebPIE encodes the materialization process into a set of MapReduce programs
while Inferray executes each rule on a dedicated thread. Our experiments show
that \sys{} outperforms all these engines in a single-core scenario. This
motivates further research on parallelizing TG-based materialization.


A second approach is to reduce the number of logically redundant facts by
appropriately ordering the rules. In~\cite{webpie}, the authors describe a rule
ordering that is optimal \emph{only} for a fixed set of RDFS rules. In contrast,
we focus on generic programs. ChaseFun~\cite{chasefun} proposes a new rule
ordering technique that focuses on \emph{equality generating dependencies}.
Hence, it is orthogonal to our approach. In a similar spirit, the rewriting
technique from \cite{motik1} targets transitive and symmetric rules. \sys{}
applies this technique by default to improve the performance, but our
experiments show it outperforms the state of the art even without this
optimization.

To optimize the execution of the rules themselves, most chase engines rely on
external DBMSs or employ state of the art query execution algorithms: LLunatic
\cite{Geerts2014b}, PDQ and ChaseFun run on top of PostgreSQL; RDFox and VLog
implement their own in-memory rule execution engine. However, none of these
engines can effectively reduce the instances over which rules are executed as
TGs do. Other approaches involve exploring columnar memory layouts as in VLog
and Inferray to reduce memory consumption and to guarantee sequential access
and efficient sort-merge join inference.

Orthogonal to the above is the work in \cite{vadalog}, which introduces a new
chase variant for materializing KBs of warded Datalog programs.  Warded Datalog
is a class of programs not admiring a finite model for any base instance.
The variant works as the restricted chase does but replaces homomorphism with
isomorphism checks.  As a result, the computed models become bigger.  An
implementation of the warded chase is also introduced in \cite{vadalog} which
focuses on decreasing the cost of isomorphism checks.  The warded chase
implementation does not apply any techniques to detect redundancies in the
offline fashion as we do for linear rules, or to reduce the execution cost of
Datalog rules as we do in Section~\ref{section:online}.


We now turn our attention to the applications of materialization in goal-driven
query answering.  Two well-known database techniques that use materialization as
a tool for goal-driven query answering are \emph{magic sets} and
\emph{subsumptive tabling}
\cite{DBLP:conf/pods/BancilhonMSU86,DBLP:journals/jlp/BeeriR91,10.1145/1989323.1989393,Sereni2008}.
The advantage of these techniques over the \emph{query rewriting} ones, which
are not based on materialization, e.g.,
\cite{DBLP:journals/semweb/CalvaneseCKKLRR17,DBLP:journals/tods/GottlobOP14,DBLP:conf/ruleml/BagetLMRS15},
is the full support of Datalog. The query rewriting techniques can support
Datalog of bounded recursion only. Beyond Datalog, materialization-based
techniques have been recently proposed for goal-driven query answering over KBs
with equality \cite{AAAI1816927}, as well as for probabilistic KBs
\cite{DBLP:conf/aaai/TsamouraGK20}, leading in both cases to significant
improvements in terms of runtime and memory consumption.  The above
automatically turns TGs to a very powerful tool to also support query-driven
knowledge exploration.

TGs are different from {acyclic graphs of rule dependencies}~\cite{agrd2}: the
former contain a single node per rule while TGs do not.  }

%% file: sections/conclusions.tex
\section{Conclusion}\label{section:conclusions}

We introduced a novel approach for materializing KBs that is based on traversing acyclic graphs of rules called TGs.
Our theoretical analysis and our empirical evaluation over well-known benchmarks
show that TG-guided reasoning is a more efficient alternative to the chase,
since it effectively overcomes all of its limitations.

Future research involves studying the problem of updating TGs in response to KB updates, as well as
extending TGs to materialize distributed KBs.

%% file: appendix/experimental-results.tex
\section{Addtional experimental results}

\begin{table}
    \centering
    \small
    \subfloat[\#Triggers for VLog and GLog on the linear scenarios.]{\label{tab:linear-triggers}\input{appendix/table-linear-triggers}}
    \;\;\;
    \subfloat[Cost of optimizations.]{\label{tab:datalog-optimizations}\input{appendix/table-datalog-optimizations}}
    \caption{Additional Experimental Results}
\end{table}




\leanparagraph{Number of triggers in the linear scenarios}
Table~\ref{tab:linear-triggers} summarizes the number of triggers for the linear scenarios.
We can see that the number of triggers in GLog is often higher
than in VLog. This is due to the fact that GLog does not eliminate redundancies right at their creation.
However, these redundancies are harmless:
due to structure sharing these redundant facts are not explicitly materialized in memory
and hence, they do not slow down the runtime.

\leanparagraph{Cost of optimizations}
Table~\ref{tab:datalog-optimizations} summarizes the cost of optimizations for the Datalog scenarios.
Recall that the optimizations in Section~\ref{section:online} are not applicable to \chaseBench as the rules have existential variables.
Column \emph{m} shows the total runtime cost of $\mathsf{minDatalog}$, while
column \emph{r} shows the total runtime cost of $\mathsf{ruleExec}$.

\input{appendix/table-unopt}

\leanparagraph{Impact on rewriting on GLog}
Tables~\ref{tab:datalog-unopt}, \ref{tab:chasebench-unopt} and \ref{tab:rdfs-unopt} summarize the performance of GLog when disabling the optimization from~\cite{motik1}. To ease the presentation, we also copy the results of the competitor engines on the same benchmarks from Tables
\ref{tab:datalog}, \ref{tab:chaseBench} and \ref{tab:rdfs}.
We can see that the only scenario whose performance degrades considerably is LUBM-LE shown in Table~\ref{tab:datalog-unopt}. Even in this case though, the performance of GLog is still better than the performance of its competitors: it is twice as fast as VLog, RDFox and COM in most scenarios and more than an order of magnitude faster than RDFox and COM in Claros-L.

\input{appendix/table-multi}

\leanparagraph{Running RDFox in multiple threads}
Tables~\ref{tab:linear-multi}, \ref{tab:datalog-multi} and \ref{tab:chasebench-multi} show the runtime performance of RDFox when increasing the number of threads from 1 to 8 and 16. For completeness, we also copy the runtime of GLog using a \emph{single thread} on the same scenarios from Tables~ \ref{tab:linear}, \ref{tab:datalog} and \ref{tab:chaseBench}. We can see that the runtime of RDFox drops considerably when using 16 threads. However, it is still higher than the runtime of GLog in all cases except UOBM-L, where RDFox's runtime is 1.6s, while GLog's runtime is 2.6s. In the other scenarios, GLog is up to 7.8 times faster than RDFox (ONT-256).

%
%

%% file: appendix/table-linear-triggers.tex
    \begin{tabular}{ p{2cm} S[table-format=6,table-column-width=1.5cm]
        S[table-format=6,table-column-width=1.5cm]}
        \textbf{Scenario} & VLog & GLog \\
        \hline
        LUBM-LI & 34346 & 35093 \\
        UOBM-LI & 7625 & 6718 \\
        DBpedia-LI & 61134 & 115150 \\
        Claros-LI & 129098 & 134800 \\
        Reactome-LI & 17120 & 23218\\
    \end{tabular}

%% file: appendix/table-datalog-optimizations.tex
    \begin{tabular}{ p{1.6cm} S[table-format=1.4,table-column-width=0.8cm] S[table-format=2.2,table-column-width=0.8cm]}
        \textbf{Scenario} & m & r \\
        \hline
        LUBM-L & 0.0005 & 0.16 \\
        LUBM-LE & 0.0007 & 0.16 \\
        UOBM-L & 0.08 & 0.02 \\
        DBpedia-L & 0.05 & 0.5 \\
        Claros-L & 0.03 & 2.8 \\
        Claros-LE & 0.3 & 15.2\\
    \end{tabular}

%% file: appendix/table-unopt.tex
\begin{table*}
    \centering
    \scriptsize
    \begin{adjustbox}{max width=\textwidth}
    {
    \begin{tabular}{ p{1.5cm}
            S[table-format=4.1,table-column-width=1cm]
            S[table-format=4,table-column-width=1cm]
        || S[table-format=4.1,table-column-width=1cm]
            S[table-format=4,table-column-width=1cm]
        || S[table-format=4.1,table-column-width=1cm]
            S[table-format=4,table-column-width=1cm]
        || S[table-format=4.1,table-column-width=0.8cm]
            S[table-format=4.1,table-column-width=0.8cm]
            S[table-format=4.1,table-column-width=0.8cm]
        || S[table-format=4,table-column-width=0.8cm]
           S[table-format=4,table-column-width=0.8cm]
           S[table-format=4,table-column-width=0.8cm]}
        & \multicolumn{2}{S[table-format=1.3,table-column-width=1.5cm]}{\textbf{VLog}}
        & \multicolumn{2}{S[table-format=1.3,table-column-width=1.5cm]}{\textbf{RDFox}}
        & \multicolumn{2}{S[table-format=1.3,table-column-width=1.5cm]}{\textbf{COM}}
        & \multicolumn{3}{S[table-format=6.2,table-column-width=1.5cm]}{\textbf{GLog Runtime}}
        & \multicolumn{3}{S[table-format=6.2,table-column-width=1.5cm]}{\textbf{GLog Memory}} \\

        \textbf{Scenario}  & {Runtime} & {Memory} & {Runtime} & {Memory} & {Runtime} & {Memory} & {No opt} & {m} & {m+r} & {No opt} & {m} & {m+r} \\

        \hline
        LUBM-L & 1.5 & 324 			& 23 & 2301 		& 20.4      & 4479 	          &  2.5 & 2.2 & 1             	 			&  446 & 424 & 265 \\
        LUBM-LE & 170.5 & 2725 		& 116.6 & 3140 		& 115.9    & 3610 	          &  71.1 &	68.8 &67.7              			&  2880	& 2688	& 2695 \\
        UOBM-L & 7.3& 1021 			& 10 & 784 			& 10        & 4215 	          &  4.4 & 6.3 & 6.3               			&  506 & 590 & 590\\
        DBpedia-L & 41.6 & 827 			& 64.4 & 3290 		& 198.4    & 3878 	          &  31.4 &	32 &  31.1               			&  2335	& 2319 & 2313 \\
        Claros-L & 431 & 3170 			& 2512 & 5491 		& 2373.0  & 6453 	          &  128.6 & 125.6 &	126.7               		&  5954 & 5957	& 5958\\
        Claros-LE & 2771.8 & 11895 		& * & * 				& *          & * 	          	    &  1104.3	& 1094 & 1106.3  		   	&  48246 & 48251 & 48223 \\
    \end{tabular} }
    \end{adjustbox}
    \vspace{0.2cm}
    \caption{{Datalog scenarios. GLog is ran without the optimization from~\cite{motik1}. Time is in sec and memory in MB.}}
    \label{tab:datalog-unopt}

    \centering
    \scriptsize
    \begin{adjustbox}{max width=\textwidth}
    {
    \begin{tabular}{ p{1.5cm}
            S[table-format=1.1,table-column-width=1cm]
            S[table-format=4,table-column-width=1cm]
        || S[table-format=2.1,table-column-width=1cm]
            S[table-format=4,table-column-width=1cm]
        || S[table-format=2,table-column-width=1cm]
            S[table-format=4,table-column-width=1cm]
        || S[table-format=2.1,table-column-width=1cm]
            S[table-format=2.1,table-column-width=1cm]
	  }
        & \multicolumn{2}{S[table-format=1.3,table-column-width=1.5cm]}{\textbf{VLog}}
        & \multicolumn{2}{S[table-format=1.3,table-column-width=1.5cm]}{\textbf{RDFox}}
        & \multicolumn{2}{S[table-format=1.3,table-column-width=1.5cm]}{\textbf{COM}}
        & \multicolumn{2}{S[table-format=6.2,table-column-width=1.5cm]}{\textbf{GLog}}\\

        \textbf{Scenario}  & {Runtime} & {Memory} & {Runtime} & {Memory} & {Runtime} & {Memory} & {Runtime} & {Memory} \\

        \hline
        STB-128 & 0.5  & 1350 			& 13.4 & 1747 		& 10      & 5217 	          & 0.2 & 1266 \\
        ONT-256 & 2.3 & 4930 			& 49 & 3997 		& 35    	& 6340 	          & 1 & 4929\\
    \end{tabular} }
    \end{adjustbox}
    \vspace{0.2cm}
    \caption{\chaseBench scenarios. GLog is ran without the optimization from~\cite{motik1}. Time is in sec and memory in MB.}
    \label{tab:chasebench-unopt}

    \centering
    \scriptsize
    \begin{adjustbox}{max width=\textwidth}
    {
    \begin{tabular}{ p{1.5cm}
            S[table-format=1.1,table-column-width=1cm]
            S[table-format=4,table-column-width=1cm]
        || S[table-format=2.1,table-column-width=1cm]
            S[table-format=4.1,table-column-width=1cm]
        || S[table-format=2.1,table-column-width=1cm]
            S[table-format=4.1,table-column-width=1cm]
        || S[table-format=4.1,table-column-width=0.8cm]
            S[table-format=4.1,table-column-width=0.8cm]
            S[table-format=4.1,table-column-width=0.8cm]
        || S[table-format=4,table-column-width=0.8cm]
           S[table-format=4,table-column-width=0.8cm]
           S[table-format=4,table-column-width=0.8cm]
	  }
        & \multicolumn{2}{S[table-format=1.3,table-column-width=1.5cm]}{\textbf{VLog}}
        &
        \multicolumn{2}{S[table-format=6.2,table-column-width=1.5cm]}{\textbf{WebPIE}}
        & \multicolumn{2}{S[table-format=6.2,table-column-width=1.5cm]}{\textbf{Inferray}}
        & \multicolumn{3}{S[table-format=6.2,table-column-width=1.5cm]}{\textbf{GLog Runtime}}
        & \multicolumn{3}{S[table-format=6.2,table-column-width=1.5cm]}{\textbf{GLog Memory}}\\

        \textbf{Scenario}  & {Runtime} & {Memory} & {Runtime} & {Memory} & {Runtime} & {Memory} & {No opt} & {m} & {m+r} & {No opt} & {m} & {m+r} \\

        \hline
        LUBM & 0.1  & 189 				& 353  & 200  			& 23 & 2000 			& 0.4 &	0.4 & 0.3  		& 186 & 187	& 181 \\
        YAGO & 163  & 3192 			& 808  & 200 			& 116.6 & 14000 		& 20  & 23 & 25 			& 1438 & 1602 &	1600 \\
    \end{tabular} }
    \end{adjustbox}
    \vspace{0.2cm}
    \caption{{RDFS scenarios. GLog is ran without the optimization from~\cite{motik1}. Time is in sec and memory in MB.}}
    \label{tab:rdfs-unopt}
\end{table*}

%% file: appendix/table-multi.tex
\begin{table*}
    \centering
    \scriptsize
    \begin{adjustbox}{max width=\textwidth}
    {
    \begin{tabular}{ p{1.5cm}
        S[table-format=2.1,table-column-width=2.5cm]
        S[table-format=2.1,table-column-width=2.5cm]
        S[table-format=2.1,table-column-width=2.5cm]
        S[table-format=2.1,table-column-width=2.5cm]
        || S[table-format=1.3,table-column-width=1.3cm]
        S[table-format=1.1,table-column-width=1.3cm]
	}
        & \multicolumn{4}{S[table-format=1.3,table-column-width=2.5cm]}{\textbf{RDFox}}
        & \multicolumn{2}{S[table-format=1.3,table-column-width=1.5cm]}{\textbf{GLog Runtime}} \\

        \textbf{Scenario}  & {Runtime (1 thread)} & {Runtime (8 threads)} & {Runtime (16 threads)}  & {Runtime (32 threads)} & {w/o cleaning}  & {w/ cleaning} \\
        \hline
        LUBM-LI & 22  & 4.7 & 3.7 & 4.3					& 0.010 & 1.1 \\
        UOBM-LI & 3.9  & 0.9 &	0.8 & 1.6  				& 0.012 & 0.2 \\
        DBpedia-LI & 44.1 & 12.8 & 10.8 & 14.7 			& 0.980 & 4.5 \\
        Claros-LI & 78.4 & 16.1 & 12 & 14				& 0.051 & 4.8 \\
        React.-LI & 12.7 & 2.8 & 2.3 & 2.9				& 0.131 & 0.9 \\
    \end{tabular} }
    \end{adjustbox}
        \caption{Linear scenarios. RDFox is ran in one, eight, 16 threads and 32 threads. GLog is ran in a single thread. Time in sec.}
    \label{tab:linear-multi}

    \centering
    \scriptsize
    \begin{adjustbox}{max width=\textwidth}
    {
    \begin{tabular}{ p{1.5cm}
            S[table-format=4.1,table-column-width=2.5cm]
            S[table-format=3.1,table-column-width=2.5cm]
            S[table-format=3.1,table-column-width=2.5cm]
            S[table-format=3.1,table-column-width=2.5cm]
         || S[table-format=4.1,table-column-width=0.8cm]
           S[table-format=4.1,table-column-width=0.8cm]
           S[table-format=4.1,table-column-width=0.8cm]
        }
        & \multicolumn{4}{S[table-format=1.3,table-column-width=2.5cm]}{\textbf{RDFox}}
        & \multicolumn{3}{S[table-format=6.2,table-column-width=1.5cm]}{\textbf{GLog Runtime}}\\

        \textbf{Scenario}  & {Runtime (1 thread)} & {Runtime (8 threads)} & {Runtime (16 threads)} & {Runtime (32 threads)} & {No opt} & {m} & {m+r} \\

        \hline
        LUBM-L 	& 23 & 4.8 & 3.8 	& 4.3				&2.4 & 2.2 & 1.0                     \\
        LUBM-LE 	& 116.6 & 20.9 & 16.2 & 17.2			& 17.3 & 17.2 & 16.1               \\
        UOBM-L 	& 10 & 2 & 1.6 & 2.4					& 2.6 & 2.4 & 2.6                    \\
        DBpedia-L 	& 64.4 & 29 & 23.9 & 34				& 20 & 19 & 19                       \\
        Claros-L 	& 2512 & 296 & 171.1 & 244.9		& 122 & 118.3 & 119               \\
        Claros-LE   & * & * & * & * 						& 1040.8 & 1012.2 & 1053.9    \\
    \end{tabular} }
    \end{adjustbox}
    \vspace{0.2cm}
    \caption{{Datalog scenarios. RDFox is ran in one, eight, 16 threads and 32 threads. GLog is ran in a single thread. Time in sec. * denotes runtime exception.}}
    \label{tab:datalog-multi}

    \centering
    \scriptsize
    \begin{adjustbox}{max width=\textwidth}
    {
    \begin{tabular}{ p{1.5cm}
            S[table-format=2.1,table-column-width=2.5cm]
            S[table-format=2.1,table-column-width=2.5cm]
            S[table-format=1.1,table-column-width=2.5cm]
            S[table-format=1.1,table-column-width=2.5cm]
        || S[table-format=1.1,table-column-width=1cm]
	  }
        & \multicolumn{4}{S[table-format=1.3,table-column-width=2.5cm]}{\textbf{RDFox}}
        & \multicolumn{1}{S[table-format=6.2,table-column-width=1.5cm]}{\textbf{GLog}~{no opt}}\\

        \textbf{Scenario}  & {Runtime (1 thread)} & {Runtime (8 threads)} & {Runtime (16 threads)} & {Runtime (32 threads)} & {Runtime} \\

        \hline
        STB-128 & 13.4 & 2.9 & 2.3 & 3.1	          & 0.2 \\
        ONT-256 & 49 & 10 & 7.8 & 10	          & 1 \\
    \end{tabular} }
    \end{adjustbox}
    \vspace{0.2cm}
    \caption{\chaseBench scenarios. RDFox is ran in 1, 8, 16 threads and 32 threads. GLog is ran in a single thread. Time in sec.}
    \label{tab:chasebench-multi}
\end{table*}

%% file: appendix/definitions.tex
We provide some definitions that will be useful for the proofs in the next section. 

For a KB \K where \Chase{\K} is defined,
the depth $\Dep{t}$ of a term $t$ in \Chase{\K} is defined as follows:
if ${t \in \Consts}$, then ${\Dep{t} = 1}$;
otherwise, if $t$ is a null of the form $n_{\ERule, \Hom, z}$, then ${\Dep{t} = {\Max{\Dep{t_1}, \dots, \Dep{t_n}} + 1}}$,
where ${\{t_1, \ldots, t_n\}}$ are all terms in the range of \Hom.

Next, we recapitulate the definitions of some known classes of programs. 
\begin{definition}\label{definition:rule-classes}
Consider a program \P and some ${k \geq 0}$.
\begin{compactitem}
	\item \P is \emph{Finite Expansion Set} (\Fes), if for each base instance $\BI$, the KB ${( \P, \BI )}$ has a terminating chase.
	\item \P is $k$-\emph{Term Depth Bounded} ($k$-\Tdb), if for each base instance $\BI$, each ${i \geq 0}$ and each term $t$ in ${\Step{i}{\P, \BI}}$, ${\Depth{t} \leq k}$.
	\P is \Tdb, if it is $k$-\Tdb.
	\item \P is \emph{Finite Order Rewritable} (\Fus), if, for each BCQ $Q$, there is a union of BCQs (UBCQs) $Q'$ such that, for each base instance $B$, we have that ${(\P, B) \models Q}$ iff ${B \models Q'}$.
\end{compactitem}
\end{definition} 

%% file: appendix/fulltg.tex
	For a program $P$, we refer to the graph computed by applying the base and the inductive steps from Section~\ref{section:trigger-graph}
	as the \emph{level-$k$ full} EG for $P$, and denote it as $\full{P}{k}$. 
	Below, we show that reasoning via a level-$k$ 
	full EG produces logically equivalent facts with the $k$-th round of the chase when the chase is applied in a breadth-first fashion:

	\begin{restatable}{theorem}{thmfulltriggergraph}\label{lemma:kboundedness:chase-equivalence}
	        For a program $P$, a base instance $B$ and a ${k \geq 0}$, ${\Step{k}{P, B} \equiv \full{P}{k}(B)}$.
	\end{restatable}
    
    	\begin{proof}
	  Let ${P = \{r_1,\dots,r_n\}}$.
	  Let ${\full{P}{k} = G^k = (V^k, E^k)}$ and let ${\full{P}{k+1} = G^{k+1} = (V^{k+1}, E^{k+1})}$.
	  Let ${I^k = \Step{k}{( P, B )}}$ and let ${J^k = G^k(B)}$.

        ($\Leftarrow$)
        To prove the claim, we show the following property, for each ${k \geq 0}$:
	  \begin{compactitem}
        	\item $\psi$. there exists a homomorphism ${g^k: J^k \rightarrow I^k}$.
	  \end{compactitem}

	  For ${k=0}$, $\psi$ holds, since ${I^0 = G^0(B) = B}$.
	  For ${k+1}$ and assuming that $\psi$ holds for $k$, the proof proceeds as follows.
	  Let ${v_1^{k+1}, \dots, v_m^{k+1}}$ be all nodes in $V^{k+1}$ of depth ${k+1}$.
	  For each ${1 \leq \iota \leq m}$, let ${\rul{v_{\iota}^{k+1}} = r_{\iota}}$ and
	  let ${v_{\iota}^{k+1}(B) = \{F_{\iota,1}, \dots, F_{\iota,n_{\iota}}\}}$.
	  Since each rule has a single atom in its head,
	  it follows from Definition~\ref{definition:guided-chase} that for each ${1 \leq \iota \leq m}$
	  and each ${1 \leq \kappa \leq n_{\iota}}$,
	  there exists a homomorphism ${h_{\iota,\kappa}}$,
	  such that ${h_{{\iota,\kappa}_s}(\head{r_{\iota}}) = F_{\iota,\kappa}}$.
	  We distinguish the cases, for each rule $r_{\iota}$, for ${1 \leq \iota \leq m}$:
	  \begin{compactitem}
		\item $r_{\iota}$ is an extensional rule. Hence, for each each ${1 \leq \kappa \leq n_{\iota}}$,
	  	${h_{\iota,\kappa}}$ is a homomorphism from $\body{r_{\iota}}$ into $B$.

		\item $r_{\iota}$ is not an extensional rule.
		WLOG, assume that $r_{\iota}$ comprises only \IP-atoms in its body.
		Since $G^{k+1}$ is a full EG,
		it follows that for each ${1 \leq \lambda \leq e_{\iota}}$,
		there exists an edge ${u_{\iota,\lambda} \rightarrow_{\iota, \lambda} v_{\iota}^{k+1}}$ in $E^{k+1}$.
		Due to the above, and due to Definition~\ref{definition:guided-chase}, it follows that
 		for each ${1 \leq \kappa \leq n_{\iota}}$,
	  	${h_{\iota,\kappa}}$ is a homomorphism from $\body{r_{\iota}}$
		into ${\bigcup \limits_{\lambda}^{e_{\iota}} u_{\iota,\lambda}(B)}$.
	  \end{compactitem}
	  Let ${N^k = \bigcup \limits_{\iota}^{m} \bigcup \limits_{\kappa}^{n_{\iota}} h_{{\iota,\kappa}_s}(\head{r_{\iota}})}$.
	  We further distinguish the cases:
	  \begin{compactitem}
		\item for each ${1 \leq \iota \leq m}$ and each ${1 \leq \kappa \leq n_{\iota}}$,
		${F_{\iota,\kappa} \in G^k(B)}$.
		Then $\psi$ trivially holds.

		\item there exists ${1 \leq \iota \leq m}$ and ${1 \leq \kappa \leq n_{\iota}}$, such that ${F_{\iota,\kappa} \not\in G^k(B)}$.
		Since $\psi$ holds for $k$ and due to $g^k$, it follows that
		for each ${1 \leq \iota \leq m}$
	 	and each ${1 \leq \kappa \leq n_{\iota}}$,
		there exists a homomorphism ${\chi_{\iota,\kappa} = h_{\iota,\kappa} \circ g^k}$
		from $\body{r_\iota}$ into $I^k$.
		Due to the above, for each ${1 \leq \iota \leq m}$
	 	and each ${1 \leq \kappa \leq n_{\iota}}$,
		there also exists a homomorphism
		$\omega_{\iota,\kappa}$
		from ${h_{{\iota,\kappa}_s}(\head{r_{\iota}})}$ into ${\chi_{{\iota,\kappa}_s}(\head{r_{\iota}})}$,
		mapping each $h_{{\iota,\kappa}_s}(n_z)$ to $\chi_{{\iota,\kappa}_s}(n'_z)$,
		where ${h_{{\iota,\kappa}_s}(z) = n_z}$ and ${\chi_{{\iota,\kappa}_s}(z) = n'_z}$,
		for each existentially quantified variable $z$ occurring in $r$.
		Due to the above and since
		for each ${1 \leq \iota \leq m}$
	 	and each ${1 \leq \kappa \leq n_{\iota}}$,
		${h_{{\iota,\kappa}_s}(c) = g^k(c)}$,
		there also exists a homomorphism
		from $N^k$ into ${M^k = \bigcup \limits_{\iota}^{m} \bigcup \limits_{\kappa}^{n_{\iota}} \chi_{{\iota,\kappa}_s}(\head{r_{\iota}})}$ and
		hence from ${I^k \cup N^k}$ into ${J^k \cup M^k}$.
		The above shows that $\psi$ holds for $k+1$.
	  \end{compactitem}

        ($\Rightarrow$)
        To prove the claim, we show the following property, for each ${k \geq 0}$:
 	  \begin{compactitem}
	        \item $\phi$. there exists a homomorphism ${g^k: I^k \rightarrow J^k}$
	        mapping each ${F \in I^k}$ to some fact ${F' \in J^k}$ of the same depth.
	  \end{compactitem}
        For ${k=0}$, $\phi$ holds, since ${I^0 = G^0(B) = B}$.
        For ${k+1}$ and assuming that $\phi$ holds for $k$ the proof proceeds as follows.
	  For each ${1 \leq i \leq n}$ and each ${1 \leq j \leq n_i}$,
        let $h_{i,j}$ be the $j$-th homomorphism from
        the body of rule $r_i$ into $I^k$, where
        ${h_{i,j}(\body{r_i})}$ comprises at least one fact of depth $k$.
        Due to the inductive hypothesis, we know that for each
        ${1 \leq i \leq n}$ and ${1 \leq j \leq n_i}$,
        there exists a homomorphism ${\chi_{i,j} = h_{i,j} \circ g^k}$
        from $\body{r_i}$ into $J^k$.
	 We distinguish the cases, for each rule $r_i$, for ${1 \leq i \leq n}$:
	 \begin{compactitem}
		\item $r_{\iota}$ is an extensional rule.
		 Hence, ${I^k = B}$.

		\item $r_{\iota}$ is not an extensional rule.
		WLOG, assume that $r_{\iota}$ comprises only \IP-atoms in its body.
		  Let ${u_{i,j}^1,\dots,u_{i,j}^{|\body{r_i}|}}$
		  be the nodes in $V^k$, such that
		  for each ${1 \leq j \leq n_i}$ and
		  each ${1 \leq l \leq |\body{r_i}|}$,
		  the $l$-th fact in ${\chi_{i,j}(\body{r_i})}$
		  belongs to ${u_{i,j}^l(B)}$.
		  Since $\phi$ holds for $k$, it follows that
 		  for each ${1 \leq j \leq n_i}$, some fact in
		  ${\chi_{i,j}(\body{r_i})}$ is of depth $k$.
		  Hence some node in
		  ${u_{i,j}^1,\dots,u_{i,j}^{|\body{r_i}|}}$ is of depth $k$.
		  Since $G^{k+1}$ is a full EG for $P$, it follows that
		  for each ${1 \leq j \leq n_i}$ and
		  each ${1 \leq l \leq |\body{r_i}|}$,
		  ${u_{i,j}^l \rightarrow_j v \in E^{k+1}}$,
	        where ${\rul{v} = r_i}$.
	 \end{compactitem}

        Due to the above and due to Definition~\ref{definition:guided-chase}, it follows that
	  for each ${1 \leq i \leq n}$ and each ${1 \leq j \leq n_i}$, we have
 	  ${\chi_{{i,j}_s}(\head{r_i}) \in G^{k+1}(B)}$.
	  Since for each ${1 \leq i \leq n}$ and each ${1 \leq j \leq n_i}$,
        there exists a homomorphism from
        $g^k$ from
        ${h_{i,j}(\body{r_i})}$ into ${\chi_{i,j}(\body{r_i})}$,
	  it follows that for each
        ${1 \leq i \leq n}$ and each ${1 \leq j \leq n_i}$,
        there also exists a homomorphism
        $\omega_{i,j}$ from
        ${h_{{i,j}_s}(\head{r_i})}$ into ${\chi_{{i,j}_s}(\head{r_i})}$,
        mapping each $h_{{i,j}_s}(n_z)$ to $\chi_{{i,j}_s}(n'_z)$,
	  where ${h_{{i,j}_s}(z) = n_z}$ and ${\chi_{{i,j}_s}(z) = n'_z}$,
	  for each existentially quantified variable $z$ occurring in $r$.
        Due to the above and since for each ${1 \leq i \leq n}$ and each ${1 \leq j \leq n_i}$,
	  ${h_{{i,j}_s}(c) = g^k(c)}$,
	  there also exists a homomorphism
        from ${N^k = \bigcup \limits_{i}^{n} \bigcup \limits^{n_i}_{j}h_{{i,j}_s}(\head{r_i})}$
        into ${M^k = \bigcup \limits_{i}^{n} \bigcup \limits^{n_i}_{j} \chi_{{i,j}_s}(\head{r_i})}$ and
        hence from ${I^k \cup N^k}$ into ${J^k \cup M^k}$.
        The above shows that the induction holds for $k+1$.
    \end{proof}

%% file: appendix/classes.tex
\thmrelationship*
\begin{proof}
	The proof is based on the proofs of Theorem \ref{theorem:ftg-bdd} and Theorem \ref{theorem:bdd-tdb-fus}.
\end{proof}

%% file: appendix/ftgbdd.tex
\begin{restatable}{theorem}{thmftgbdd}\label{theorem:ftg-bdd}
	\P is \Ftg iff it is \Bdd.
\end{restatable}
\begin{proof}
$\implies$
If $\P$ is \Ftg, then there is some finite TG $\TG = (V, E)$ for this program.
We proceed to show that \P is \KBdd with $k$ the depth of $\TG$.
More precisely, we show by contradiction that $\Step{k}{ \P, \BI } \models \Step{k+1}{ \P, \BI }$ for any given base instance \BI.
\begin{enumerate}
\item Suppose for a contradiction that $\Step{k}{\K} \not \models \Step{k+1}{\K}$ with $\K = (\P, \BI)$.
\item By the definition of the standard chase, $\Step{k}{\K} \subseteq \Step{k+1}{\K}$.
\item By (1) and (2), there is some BCQ \Query such that $\Step{k}{\K} \not\models \Query$ and $\Step{k+1}{\K} \models \Query$.
\item We can show via induction that $\TG^i(\BI) \subseteq \Step{i}{\K}$ for all $i \geq 0$.
\item By (4), $\TG^k(\BI) \subseteq \Step{k}{\K}$.
\item By (3) and (5), $\TG(\BI) \not\models \Query$.
\item By (3), $\K \models \Query$.
\item By (6) and (7), the graph \TG is not a TG for \P.
\end{enumerate}

$\impliedby$
Since $\P$ is \Bdd, we have that $\P$ is $k\text{-}\Bdd$ for some $k \geq 0$.
Therefore, the graph $\full{P}{k}$ is a TG for \P by Theorem~\ref{lemma:kboundedness:chase-equivalence} and the program $\P$ is $\Ftg$.
\end{proof}

%% file: appendix/bddtdbfus.tex
\begin{restatable}{theorem}{thmbddtdbfus}\label{theorem:bdd-tdb-fus}
	\P is \Tdb and \Fus iff it is \Bdd.
\end{restatable}
\begin{proof}
$\implies$
\begin{enumerate}
\item Assume that $P$ is (a) \Fus and (b) \Tdb.
\item Let $h$ be a homomorphism that maps every $t \in \Nulls \cup \Vars$ to a fresh $c_t \in \Constants$ unique for $t$.
\item For all facts $\varphi$ that can be defined over some $s \in \Predicates$ in $P$, we introduce the following notions.
\begin{enumerate}[(a)]
\item Let $\Rw_\varphi$ be some (arbitrarily chosen) rewriting for the BCQ $\varphi$ with respect to $P$.
Note that, such a rewriting must exist by (1.a).
\item Let $k_{\varphi}$ be the smallest number such that $\Step{k_{\varphi}}{P, h(\beta)} \models \Step{k_{\varphi}+1}{P, h(\beta)}$ for every disjunct $\beta$ in the rewriting $\Rw_{\varphi}$.
Note that, such a number must exist by (1.b).
\end{enumerate}
\item For all $s \in \Predicates$, let $k_s$ be the smallest number such that $k_s > k_\varphi$ for all facts $\varphi$ that can be defined over the predicate $s$.
Note that, the number $k_s$ is well-defined despite the fact that we can define infinitely many different facts over any given predicate.
This is because $k_{s(t_1, \ldots, t_n)} = k_{s(u_1, \ldots, u_n)}$ if we have that there is a bijective function mapping $t_i$ to $u_i$ for all $1 \leq i \leq n$.
\item Let $k_P$ be the smallest number such that $k_P > k_s + 1$ for all $s \in \Predicates$ in $P$.
\item Consider some fact $\varphi = s(t_1, \ldots, t_n)$, some base instance $B$, and some $i \geq 0$.
We show that, if the terms $t_1, \ldots, t_n$ are in \Step{i}{P, B} and $\varphi \in \Chase{P, B}$, then $\varphi \in \Step{i + k_P - 1}{P, B}$.
\begin{enumerate}[a.]
\item $h(\varphi) \in \Chase{(P, B')}$ with $B' = h(\Step{i}{P, B})$.
\item By (a): $B' \models \Rw_{h(\varphi)}$ where $\Rw_{h(\varphi)}$ is a UBCQ of the form $\exists \vec{x}_1 . \beta_1 \vee \ldots \vee \exists \vec{x}_n . \beta_n$.
\item By (b): $B' \models \exists \vec{x}_k . \beta_k$ for some $1 \leq k \leq n$.
\item By (c): there is a homomorphism such that $h'(\beta_k) \subseteq B'$.
\item By (d): $h'(\varphi) \in \Chase{(P, h(\beta_k))}$.
\item By (5) and (e): $\Step{k_P-1}{P, h'(\beta_k)} \supseteq \Chase{P, h'(\beta_k)}$.
\item By (e) and (f): $h'(\varphi) \in \Step{k_P-1}{P, h'(\beta_k)}$.
\item By (d) and (g): $h'(\varphi) \in \Step{k_P-1}{P, B'}$.
\item By (h): $\varphi \in \Step{i+ k_P-  1}{P, B}$.
\end{enumerate}
\item For a fact $\varphi = s(t_1, \ldots, t_n)$, let $\Dep{\varphi} = \textit{max}_{1 \leq i \leq n}(\Dep{t_i})$.
\item By (6) and (7): We show via induction that, for all $d \geq 1$, the set $\Step{d \cdot k_P}{P, F}$ contains all of the facts $\varphi \in \Chase{P, F}$ with $\Dep{\varphi} \leq d$.
\begin{itemize}
\item Base case:
The set \Step{0}{P, F} contains all terms of depth $1$ (i.e., all constants) that occur in \Chase{P, F}.
Hence, by (6), the set \Step{k_P}{P, F} contains every $\varphi \in \Chase{P, F}$ with $\Dep{\varphi} = 1$.
\item Inductive step:
Let $i \geq 1$.
Then, \Step{(i-1) \cdot k_P}{P, F} contains all $\varphi \in \Chase{P, F}$ with $\Dep{\varphi} = i-1$.
Hence, the set $\Step{(i-1) \cdot k_P + 1}{P, F}$ contains all $t \in \Terms$ in \Chase{P, F} with $\Dep{\varphi} = i$.
By (6), the set $\Step{i \cdot k_P}{P, F}$ contains all $\varphi \in \Chase{P, F}$ with $\Dep{\varphi} = i$.
\end{itemize}
\item By (1.b): There is some $k_d \geq 0$ that depends only on $P$ such that, for every term in $t$ occurring in $\Chase{(P, F)}$, the depth $t$ is equal or smaller than $k_d$.
\item By (8) and (9): $P$ is $(k_d \cdot k_P)$-\Bdd.
Note that, neither $k_d$ nor $k_P$ depend on the set of facts $F$.
\end{enumerate}


$\impliedby$
Since $\P \in \Bdd$, $\P \in \KBdd$ for some $k \geq 0$.
Via induction on $i \in \{0, \ldots, k\}$, we can show that all terms in \Step{i}{\P, \BI} are of depth $i$ or smaller for any instance \BI, and hence, $\P \in \KTdb$.
Note that $\Bdd \subseteq \Fus$ has been shown in \cite{DBLP:conf/dlog/LeclereMU16}.
\end{proof}

%% file: appendix/ftgundecidability.tex
\begin{theorem} \label{theorem:undecidability-finite-trigger-graph}
	The language of all programs that admit a finite TG is undecidable.
\end{theorem}
\begin{proof}
This follows from the fact that \Ftg decidability implies decidability of \Fus
for Datalog programs, which is undecidable \cite{DBLP:conf/adbis/GottlobOP11}.
\end{proof}

%% file: appendix/existence.tex
\thmTGexistence*

\begin{proof}
     Recall that we denote by $\full{P}{k}$ the level-$k$ full EG for a program $P$. 
     We can stop the expansion of the level-$k$ full EG for a program $P$
     when ${\full{P}{k-1}(B) \models \full{P}{k}(B)}$ holds. 
     The above, along with Theorem~\ref{lemma:kboundedness:chase-equivalence} 
     and the fact that whenever ${\Step{k-1}{P, B} \models \Step{k}{P, B}}$ holds for some ${k \geq 1}$, 
     then the KB ${(P,B)}$ admits a finite model (see \cite{k-boundedness}) conclude the proof of Theorem~\ref{theorem:TGexistence}.
\end{proof}

%% file: appendix/tglinear.tex
\thmtriggerGraphLinear*

\begin{proof}
In order to show that the EG ${G = (V,E)}$ computed by Algorithm~\ref{alg:Trigger-Graph-Computation-Linear}
is a TG for $P$, it suffices to show that
$\Step{\infty}{P, B}$ is logically equivalent to $G(B)$.
The proof makes use of Propositions~\ref{proposition:linear:chase-separability} and \ref{proposition:linear:node-facts},
as well as of Lemma~\ref{lemma:linear:node-existence}.
Note that Propositions~\ref{proposition:linear:chase-separability} is a known result and, hence, we skip its proof, while Proposition~\ref{proposition:linear:node-facts}
directly follows from Definition~\ref{definition:guided-chase}.

\begin{proposition}\label{proposition:linear:chase-separability}
	For each linear program $P$ and each base instance $B$, the following holds:
	\begin{align}
		\Step{\infty}{P, B} \equiv \bigcup \limits_{F \in B} \Step{\infty}{P,\{F\}}										\label{eq:linear:chase-separability}
	\end{align}
\end{proposition}

\begin{proposition}\label{proposition:linear:node-facts}
	For each linear TG $G$, each node ${v \in G}$ and each base instance $B$, the following holds:
	\begin{align}
			v(B)  = \bigcup \limits_{F \in B} u(\{F\})
	\end{align}
\end{proposition}

\begin{lemma}\label{lemma:linear:node-existence}
	For each fact ${f \in \mathcal{H}(P)}$, ${\Gamma(\{f\}) \equiv \Step{\infty}{P,\{f\}}}$, where $\Gamma$ is the EG computed in lines \ref{alg:Trigger-Graph-Computation-Linear:tgraph:start}--\ref{alg:Trigger-Graph-Computation-Linear:tgraph:end} for fact $f$.
\end{lemma}
\begin{proof}
	($\Rightarrow$)
	We want to show that for each fact ${f \in \mathcal{H}(P)}$,
	there exists a homomorphism from ${\Gamma(\{f\})}$ into ${\Step{\infty}{P, \{f\}}}$.
	Due to Algorithm~\ref{alg:Trigger-Graph-Computation-Linear}, we know that
	$\Gamma$ is a graph of the form
	\begin{align}
		\left( \bigcup \limits_{j=1}^n u_j, \bigcup \limits_{j=1}^{n-1} \{ u_j \rightarrow_1 u_{j+1} \} \right)
	\end{align}
	where ${u_j \in V}$,  for each ${0 \leq i \leq n}$; ${u_j \rightarrow_1 u_{j+1} \in E}$, for each ${0 \leq i < n}$;
	and ${u_n = v}$.
	To prove this direction, we will show that the following property holds, for each ${0 \leq i \leq n}$:
	\begin{compactitem}
		\item $\phi_1$. there exists a homomorphism $h^i$ from ${\Gamma_{\preceq u_i}(\{f\})}$ into
		${\Step{\infty}{P, \{f\}}}$.
	\end{compactitem}
	For ${i=0}$, since ${\Gamma_{\preceq v_i}}$ is the empty graph and hence
	${\Gamma_{\preceq u_i}(B) = B}$ by Definition~\ref{definition:guided-chase}, it follows that $\phi_1$ holds.
	For ${i+1}$ and assuming that $\phi_1$ holds for $i$ the proof proceeds as follows.
	Since $\phi_1$ holds for $i$, we know that there exists a homomorphism $h^i$ from
	${\Gamma_{\preceq u_i}(\{f\})}$ into ${\Step{\infty}{P, \{f\}}}$.
	If ${u_{i+1}(\{f\}) = \emptyset}$, then $\phi_1$ trivially holds for ${i+1}$.
	Hence, we will consider the case where ${v_{i+1}(\{f\}) \neq \emptyset}$.
	Since $v_{i+1}$ is a child of $v_i$ and since $v_{i+1}$ is associated with some
	rule ${r_{i+1} \in P}$, it follows that
	there exists a homomorphism $g$ from $\body{r_{i+1}}$ into
	${v_i(\{f\})}$ and ${v_{i+1}(\{f\}) = g_s(\head{r_{i+1}})}$.
	Due to $g$ and due to $h^i$, we know that there exists a homomorphism
	${\psi = g \circ h^i}$ from $\body{r_{i+1}}$ into ${\Step{\infty}{P, \{f\}}}$ and, hence,
	a homomorphism $\omega$ from ${g_s(\head{r_{i+1}})}$ into ${\psi_s(\head{r_{i+1}})}$
	mapping each value $c$ occurring in $\domain{g}$ into ${(g \circ h^i)(c)}$ and
	$n_z$ into $n'_z$, for each existentially quantified variable $z$ of $r_{i+1}$,  
	where ${g(z) = n_z}$ and ${\psi(z) = n'_z}$.
	We distinguish the cases:
	\begin{compactitem}
		\item ${\psi_s(\head{r_{i+1}}) \in \Step{\infty}{P, \{f\}}}$.
		Due to $h^i$ and due to $\omega$, it follows that
		${h^{i+1} = h^i \cup \omega}$ is a homomorphism from
		${\Gamma_{\preceq v_{i+1}}(\{f\})}$ into ${\Step{\infty}{P, \{f\}}}$.

		\item ${\psi_s(\head{r_{i+1}}) \not \in \Step{\infty}{P, \{f\}}}$.
		Then ${\Step{\infty}{P, \{f\}} \models \psi_s(\head{r_{i+1}})}$ holds and hence,
		there exists a homomorphism $\theta$ from
		${\psi_s(\head{r_{i+1}})}$ into ${\Step{\infty}{P, \{f\}}}$.
		Due to $h^i$, due to $\omega$ and due to $\theta$, we can see that
		${h^{i+1} = (h^i \cup \omega) \circ \theta}$ is a homomorphism
		from
		${\Gamma_{\preceq v_{i+1}}(\{f\})}$ into ${\Step{\infty}{P, \{f\}}}$.
 	\end{compactitem}
 	The above shows that $\phi_1$ holds for ${i+1}$ concluding the proof of this direction.

	($\Leftarrow$)
	We want to show that for each fact ${f \in \mathcal{H}(P)}$,
	there exists a homomorphism from ${\Step{\infty}{P, \{f\}}}$ into ${\Gamma(\{f\})}$.
	We use $I^i$ to denote ${\Step{i}{P, \{f\}}}$ 
	and ${\chaseGraph^{i}(P,\{f\})}$ to denote the chase graph corresponding to ${\Step{i}{P, \{f\}}}$. 
	The proof of this direction proceeds by showing that the following properties hold, for each ${i \geq 0}$:
	\begin{compactitem}
	 	\item $\phi_2$. there exists a homomorphism ${h^i}$ from ${\Step{i}{P, \{f\}}}$ into ${\Gamma(\{f\})}$.
	 	\item $\phi_3$. for each ${f_1 \rightarrow_{r_1} f_2 \rightarrow_{r_2} \dots \rightarrow_{r_{j}} f_{j+1}}$ in 
	 	${\chaseGraph^{i}(P,\{f\})}$, a path of the form 
	 	${u_1 \rightarrow_1 \dots \rightarrow_{1} u_j}$ is in $\Gamma$, where for each ${1 \leq k \leq j}$: $u_k$ is associated with $r_k$  
	 	and there exists a homomorphism from $f_{k+1}$ into ${u_k(\{f\})}$.
	\end{compactitem}
	
	For ${i=0}$, since ${\Step{0}{P, \{f\}} = \{f\}}$ 
	and since ${f \in \Gamma(\{f\})}$ by definition, it follows that the inductive hypotheses $\phi_2$ and $\phi_3$ holds. 
	For ${i+1}$ and assuming that $\phi_2$ and $\phi_3$ hold for $i$ the proof proceeds as follows.
	Let $\Sigma_r$ be the set of all active triggers for each ${r \in P}$ in $I^i$.
	Let also 
	\begin{align}
		\Delta I = \bigcup \nolimits_{r \in P} \bigcup \nolimits_{h \in \Sigma_r} h_s(\head{r})		\label{eq:active_triggers}
	\end{align}
	
	We distinguish the following cases: 
	\begin{compactitem}	
		\item ${I^i \models I^i \cup \Delta I}$ holds.  
		Then the equivalent chase terminates and hence ${\Step{i}{P, \{f\}} = \Step{\infty}{P, \{f\}}}$.
		Since the inductive hypotheses $\phi_2$ and $\phi_3$ hold for $i$ and due to the above, it follows that the inductive hypotheses will hold for ${i+1}$.   

		\item ${I^i \models I^i \cup \Delta I}$ does not hold. Then, 
		for each rule ${r \in P}$ for which ${\Sigma_r \neq \emptyset}$ and for each ${h \in \Sigma_r}$, 
		the chase graph ${\chaseGraph^{i+1}(P,\{f\})}$ will include an edge ${h(\body{r}) \rightarrow_r h_s(\head{r})}$. 
  	      Due to the steps in lines \ref{alg:Trigger-Graph-Computation-Linear:tgraph:node:start}--\ref{alg:Trigger-Graph-Computation-Linear:tgraph:node:end}, we know that 
  	      $\Gamma$ includes a node $v$ associated with $r$ $(\ast)$.
  	      We have the following two subcases:
  	      \begin{compactitem}
  	      		\item There is no edge of the form ${f' \rightarrow_{r'} h(\body{r})}$ in ${\chaseGraph^{i+1}(P,\{f\})}$, 
	  	      for some ${r' \in P}$. Then, it follows that ${h(\body{r}) = f}$. Due to the above, due to $(\ast)$ and due to Definition~\ref{definition:guided-chase}, 
	  	      it follows that the inductive hypotheses $\phi_2$ and $\phi_3$ hold for ${i+1}$.   
	  	      
	  	      \item Otherwise. Since the inductive hypothesis $\phi_3$ holds for $i$
	  	      and due to the steps in lines \ref{alg:Trigger-Graph-Computation-Linear:tgraph:edge:start}--\ref{alg:Trigger-Graph-Computation-Linear:tgraph:edge:end}, 
	  	      it follows that an edge of the form ${v' \rightarrow_1 v}$ will be in $\Gamma$, where node $v'$ is associated with rule $r'$.  
	  	      Furthermore, due to $\phi_3$, there exists a homomorphism $g^i$ from ${h(\body{r})}$ into ${v'(\{f\})}$. 
	  	      Due to $g^i$, due to the fact that the edge ${v' \rightarrow_1 v}$ is in $\Gamma$ and due to Definition~\ref{definition:guided-chase}, 
	  	      there exists a homomorphism $g^{i+1}$ from ${h_s(\head{r})}$ into ${v(\{f\})}$.
	  	      Finally, due to the above, and since $\phi_2$ holds for $i$, it follows that
	  	      ${h^{i+1} = h^i \circ g^{i+1}}$ is a homomorphism from ${\Step{i}{P, \{f\}}}$ into $\Gamma(\{f\})$.
	  	      Hence, $\phi_2$ and $\phi_3$ hold for ${i+1}$ concluding the proof of this direction and, consequently of Lemma~\ref{lemma:linear:node-existence}.
  	      \end{compactitem}
	\end{compactitem}
\end{proof}

We are now ready to return to the proof of Theorem~\ref{theorem:triggerGraphLinear}.
Let ${\Gamma_f}$ be the EG computed in lines \ref{alg:Trigger-Graph-Computation-Linear:tgraph:start}--\ref{alg:Trigger-Graph-Computation-Linear:tgraph:end} 
for each ${f \in \mathcal{H}(P)}$.
Since Algorithm~\ref{alg:Trigger-Graph-Computation-Linear} only adds new nodes and new edges to set of nodes and the set of edges of an input EG, it follows that
\begin{align}
	V &= \bigcup_{\forall f \in \mathcal{H}(P)} \nodes{\Gamma_f} 												\label{eq:linear:nodes} \\
	E &= \bigcup_{\forall f \in \mathcal{H}(P)} \edges{\Gamma_f} 												\label{eq:linear:edges}
\end{align}

Since for each base instance $B$ of $P$ and each fact ${f' \in B}$, there exists a fact ${f \in \mathcal{H}(P)}$ and a bijective function $g$ over the constants in ${\mathcal{C}}$, such that 
${g(f') = f}$ and from Lemma~\ref{lemma:linear:node-existence},
it follows that ${\Gamma_f(\{f'\}) \equiv \Step{\infty}{P, f'}}$.
Due to the above, due to \eqref{eq:linear:nodes} and \eqref{eq:linear:edges} and since each node in $V$ has up to one incoming edge, it follows that for each base instance $B$ of $P$
\begin{align}
	\bigcup \limits_{F \in B} \Step{\infty}{P, \{F\}} \equiv \bigcup \limits_{F \in B} G(\{F\})						\label{eq:linear:tgraph1}
\end{align}
Due to Definition~\ref{definition:guided-chase} and since each node in $V$ has up to one incoming edge, it follows that for each base instance $B$ of $P$ we have 
\begin{align}
	\bigcup \limits_{F \in B} G(\{F\}) = G(B)																	\label{eq:linear:tgraph2}
\end{align}
Finally, due to Proposition~\ref{proposition:linear:chase-separability}, and due to \eqref{eq:linear:tgraph1} and \eqref{eq:linear:tgraph2},
we have ${\Step{\infty}{P, B} \equiv G(B)}$, for each base instance $B$ of $P$.
The above concludes the proof of the first part of Theorem~\ref{theorem:triggerGraphLinear}.
\end{proof}

%% file: appendix/tglinear-complexity.tex
\thmtriggerGraphLinearcomplexity*

\begin{proof}
We first show the first part of the theorem with a step-by-step argument.
\begin{enumerate}
\item Consider some program $P$.
\item The set $\mathcal{H}(P)$ is exponential in $P$.
\begin{enumerate}
\item Let \textbf{Ary} be the maximal arity of an extensional predicate occurring in $P$.
\item For an extensional predicate $p$ occurring in $P$, there are at most $\textbf{Ary}^\textbf{Ary}$ facts in $\mathcal{H}(P)$ defined over $p$.
\item Since $P$ is a linear rule set and extensional predicates may only appear in the body of a rule, we have that number of extensional predicates in $P$ is at most $\vert P \vert$.
\item By (b) and (c): there are at most $\vert P \vert \times \textbf{Ary}^\textbf{Ary}$ facts in $\mathcal{H}(P)$.
\end{enumerate}
\item For every fact $f \in \mathcal{H}(P)$, we may have to add at most $\vert \Chase{P, \{f\}} \vert^2$ nodes to the graph $\tgraphLinear(P)$.
(Note that \Chase{P, \{f\}} is defined since $P$ is \Fes.)
Therefore, this graph contains at most $\vert \Chase{P, \{f\}} \vert^4$ edges.

\item From results in \cite{DBLP:conf/icdt/LeclereMTU19}, the size of \Chase{P, \{f\}} is double exponential in $P$.
\begin{enumerate}
\item By Proposition~30 in \cite{DBLP:conf/icdt/LeclereMTU19}: if \Chase{P, \{f\}} is finite, then there exists a finite entailment tree $\mathcal{T}$ such that the set of atoms associated with $\mathcal{T}$ is a complete core.
\item In Algorithm~1 in \cite{DBLP:conf/icdt/LeclereMTU19}, the authors describe  how to construct the finite entailment tree for $P$ and $\{f\}$.
\item The complexity of this algorithm, as well as the size of the output tree, is double exponential in the input $P$ and $\{f\}$.
Note the discussion right after Algorithm~1 in \cite{DBLP:conf/icdt/LeclereMTU19}.
\end{enumerate}
\item By (2--4): the algorithm $\tgraphLinear(P)$ runs in double exponential in $P$.
\end{enumerate}

To show that the procedure $\tgraphLinear(P)$ runs in single exponential time when $(*)$ the arity of the predicates in $P$ is bounded, we can show that the size of \Chase{P, \{f\}} is (single) exponential in $P$ if $(*)$.
In fact, this claim also follows from then results in \cite{DBLP:conf/icdt/LeclereMTU19}.
Namely, if $(*)$, then entailment trees for $P$ and $f$ (as they are defined in \cite{DBLP:conf/icdt/LeclereMTU19}) are of polynomial depth because the number of ``sharing types'' for $P$ is polynomial.
Therefore, the size of these trees is exponential and so is the size of \Chase{P, \{f\}}.
\end{proof}

%% file: appendix/minlinear.tex
\lemmalinearredundant*

\begin{proof}
    ($\Rightarrow$) This direction trivially holds.

    ($\Leftarrow$) We want to show that if there exists a preserving homomorphism
    from $u(\{f\})$ into $v(\{f\})$, for each fact ${f \in \mathcal{H}(P)}$, then
    there exists a preserving homomorphism from $u(B)$ into $v(B)$, for each base instance $B$.
    The proof works by contradiction.
    Suppose that there exists a base instance $B'$, such that there does not exist
    a preserving homomorphism from $u(B')$ into $v(B')$.
    Since there does not exist
    a preserving homomorphism from $u(B')$ into $v(B')$ and
    due to Proposition~\ref{proposition:linear:node-facts},
    it follows that there does not exist a preserving homomorphism from
    ${\bigcup \nolimits_{F' \in B'} u(\{F'\})}$ into ${\bigcup \nolimits_{F' \in B'} v(\{F'\})}$.

    Next we show that there exists some ${F' \in B'}$, so that there does not
    exist a preserving homomorphism $h_{F'}$ from $u(\{F'\})$ into $v(\{F'\})$.
    The proof proceeds as follows.
    Suppose by contradiction that there exists a preserving homomorphism $h_{F'}$
    from $u(\{F'\})$ into $v(\{F'\})$, for each ${F' \in B'}$, but there does not
    exist a preserving homomorphism $h_{B'}$ from $u(\{B'\})$ into $v(\{B'\})$.
    The above assumption, will be referred to as Assumption~($\mathsf{A}_1$).
    By definition, we know that a preserving homomorphism from
    $u(\{B'\})$ into $v(\{B'\})$ maps each value $c$ either (i) to itself
    if $c$ was a schema constant or a null occurring in
    ${\left(G(B') \setminus G_{\succeq u}(B') \right) \cap G_{\succeq u}(B')}$ or (ii) to a fresh null
    occurring in a single fact from $v(\{F'\})$.
    Since for each ${F' \in B'}$, $h_{F'}$ maps
    each schema constant and each null occurring in
    ${\left(G(B') \setminus G_{\succeq u}(B') \right) \cap G_{\succeq u}(B')}$
    to itself according to Assumption~($\mathsf{A}_1$) and due to the above,
    it follows that
    there exist two facts ${F'_1,F'_2 \in B'}$ and a null $\mathsf{n}$ occurring in ${G_{\succeq u}(B') \setminus G(B')}$,
    such that ${h_{F'_1}(\mathsf{n}) = \mathsf{m}_1}$ and ${h_{F'_2}(\mathsf{n}) = \mathsf{m}_2}$, where
    ${h_{F'_j}}$ is a preserving homomorphism from  $u(\{F'_j\})$ into $v(\{F'_j\})$, for each ${1 \leq j \leq 2}$.
    However, the above leads to a contradiction, since in linear TGs  each null from $u(\{B'\})$ or $v(\{B'\})$
    occurs in only one fact. The above shows that if there
    does not exist a preserving homomorphism from
    ${\bigcup \nolimits_{F' \in B'} u(\{F'\})}$ into ${\bigcup \nolimits_{F' \in B'} v(\{F'\})}$, then
    there exists some ${F' \in B'}$, such that there does not
    exist a preserving homomorphism from $u(\{F'\})$ into $v(\{F'\})$. The proof proceeds as follows.

    Since for each ${F' \in B'}$, there exists a bijective function $g$ over the constants in ${\mathcal{C}}$
    and an instance ${f \in \mathcal{H}(P)}$, such that 
    $g(F') = f$, 
    it follows that there does not exist a preserving homomorphism from
    $u(\{f\})$ into $v(\{f\})$, for some ${f \in \mathcal{H}(P)}$.
    This leads to a contradiction.
    Hence, there exists a preserving homomorphism from $u(B)$ into $v(B)$,
    for each base instance $B$ and thus,
    Lemma~\ref{lemma:linear:redundant} holds.
\end{proof}

\thmminimizeexists*

\begin{proof}
Recall from Definition~\ref{definition:minimization} that $\mathsf{minLinear}(G)$ results from $G$ after applying the following step until reaching a fixpoint: (i) find a pair of nodes $u$ and $v$ with $u$ being dominated by $v$; (ii) remove $v$ from $\nodes{G}$; and (iii) 
add an edge ${v \rightarrow_j u’}$, for each edge ${u \rightarrow_j u’}$ from $\edges{G}$.
Let $G^i$ be the EG computed at the beginning of the $i$-th step of this iterative process with ${G^0 = G}$. 
In order to show that $\mathsf{minLinear}(G)$ is a TG for $P$, we need to show that the following property holds for each BCQ $Q$ entailed by $(P,B)$: 
\begin{align}
	G^i(B) \models Q \tag{$\ast$}
\end{align}  
We can see that $(\ast)$ holds ${i=0}$, since ${G^0 = G}$. 
For ${i+1}$ and assuming that $(\ast)$ holds for ${i \geq 0}$ we proceed as follows. 
Suppose that there exists a homomorphism $q$ from $Q$ into ${G^i(B)}$.
Furthermore, let $C_1$ and $C_2$ be two conjuncts, such that
${Q = C_1 \wedge C_2}$ and $q$ maps $C_1$ into $G_{\succeq u}(B)$
and $C_2$ into ${G(B) \setminus G_{\succeq u}(B)}$.
Due to the above, it follows that $q$ maps each variable occurring both in $C_1$ and $C_2$ either to a constant or to a null occurring in ${\left(G(B) \setminus G_{\succeq u}(B) \right) \cap G_{\succeq u}(B)}$. 

From Definition~\ref{definition:minimization} we know that (i) there exists a pair of nodes ${u,v \in \nodes{G^i}}$ with $u$ being dominated by $v$ and that 
(ii) the graph $\Gamma^{i+1}$ that results from $G^i_{\succeq v}$ after adding an edge ${v \rightarrow u’}$, for each edge ${u \rightarrow u’}$ in $\edges{G^i}$, is a subgraph of $G^{i+1}$.
Since $q$ maps each variable occurring both in $C_1$ and $C_2$ either to a constant or to a null occurring in ${\left(G(B) \setminus G_{\succeq u}(B) \right) \cap G_{\succeq u}(B)}$ it follows that 
$\ast$ holds for $i+1$ if the following holds: 

\begin{lemma}\label{lemma:linear:homomorphism}
	There exists a homomorphism $g$ from $G^i_{\succeq u}(B)$ into $\Gamma^{i+1}(B)$ so that ${g(c)=h(c)}$, for each ${c \in \domain{h}}$.   
\end{lemma}  
The proof of Lemma~\ref{lemma:linear:homomorphism} directly follows from the facts that
(i) there exists a preserving homomorphism from $u(B)$ into $v(B)$, for each base instance $B$ and (ii) all subgraphs rooted at each child of $u$ are copied below $v$.

From Lemma~\ref{lemma:linear:homomorphism} and since $\Gamma^{i+1}$ is a subgraph of $G^{i+1}$, we have that  
there exists a homomorphism $(q \circ g)$ from $Q$ into $G^{i+1}(B)$ concluding the proof of $(\ast)$ for $i+1$ and hence the proof of Theorem~\ref{theorem:triggerGraphLinear}.
\end{proof}

%% file: appendix/EG-rewriting.tex
Below, we recapitulate the notion of the answers of non-Boolean CQs on a KB. 
The answers to a CQ $Q$ on a KB ${(P,B)}$, denoted as $\mathsf{ans}(Q, P, B)$,  
is the set of tuples that are answers to each model of ${(P,B)}$, 
i.e., ${\{\vec{t} | \vec{t} \in Q(I), \text{ for each model } I \text{ of } (P,B)\}}$.  

\thmEP*

\begin{proof}

	Let $R$ be the predicate in the body of the characteristic query of $v$. 
	In order to prove Lemma~\ref{lemma:EP-rewriting}, we have to show that $\vec{t}$ is an answer to $\rew(v)$ on $B$ iff ${R(\vec{t}) \in v(B)}$. 
	The proof is based on (i) the correspondence between the rewriting process of Definition~\ref{definition:EP-rewriting} and the query rewriting algorithm from \cite{DBLP:journals/tods/GottlobOP14}, 
	called $\mathsf{XRewrite}$ and (ii) the correctness of $\mathsf{XRewrite}$.  
	In particular, the steps of the proof are as follows. 
	First, we compute a new set of rules $P^*$ by rewriting the rules associated with the nodes in ${G_{\preceq v}}$. 
	This rewriting process is described in Definition~\ref{definition:rule-rewriting}. 
	Then, we establish the relationship between the $\rew{v}$ and the rewritings computed by $\mathsf{XRewrite}$. 
	In particular, Lemma~\ref{lemma:correspondence-of-rewritings} shows that Definition~\ref{definition:EP-rewriting} and $\mathsf{XRewrite}$ 
	result in the same rewritings modulo bijective variable renaming, 
	when the latter is provided with $P^*$ and a rewriting of the characteristic query of $v$ denoted as $Q^*$. 
	A direct consequence of Lemma~\ref{lemma:correspondence-of-rewritings} is that $\mathsf{XRewrite}$ terminates when provided with $P^*$ and $Q^*$.
	In order to show our goal, we make use of the above results as well as of Lemma~\ref{lemma:correspondence-of-facts}	.  

	Below, we describe $\mathsf{XRewrite}$. 
	Given a CQ $Q$ and a set of rules $P$, $\mathsf{XRewrite}$ computes a rewriting $Q_r$ of $Q$ so that for any null-free instance $I$, 
	the answers to $Q_r$ on ${(P,I)}$ coincide with the answers to $Q$ on ${(P,I)}$.      
	We describe how $\mathsf{XRewrite}(P,Q)$ works when $P$ is Datalog. 
	At each step $i$, $\mathsf{XRewrite}$ computes a tree $\mathcal{T}^i$, where each node $\kappa$ is associated with a CQ denoted as $\qry{\kappa}$.   
	When ${i=0}$, $\mathcal{T}^0$ includes a single root node associated with $Q$.
	When ${i>0}$, then $\mathcal{T}^i$ is computed as follows: 
	for each leaf node $\kappa$ in $\nodes{\mathcal{T}^{i-1}}$, each atom $\beta$ occurring in the body of ${\qry{\kappa}}$  
	and each rule ${r \in P}$ whose head unifies with $\beta$, $\mathsf{XRewrite}$:
	\begin{enumerate}
		\item  computes a new query $Q'$ by 	(i) computing the MGU $\theta$ of ${\{\head{r},\alpha\}}$, (ii) 
      replacing $\alpha$ in the body of ${\qry{\kappa}}$ with $\body{r}$ and (iii) applying $\theta$ on the resulting query;
      		\item adds a new node $\kappa'$ in $\mathcal{T}^i$ and associates it with $Q'$; and 
      		\item adds the edge ${\kappa \xrightarrow{(\alpha,r)} \kappa'}$ to $\mathcal{T}^i$. 
	\end{enumerate}
	Notice that $\mathsf{XRewrite}$ also includes a factorization step, however, this step is only applicable in the presence of existential rules. 
	
	We now introduce some notation related to Definition~\ref{definition:EP-rewriting}. 
	We denote by ${\rew^i(v)}$ the CQ at the beginning of the $i$-th iteration of the rewriting step of Definition~\ref{definition:EP-rewriting} with 
	${\rew^0(v)}$ being equal to the characteristic query of $v$. 
	We use ${\rew^i(v) \xrightarrow{\alpha_i} \rew^{i+1}(v)}$ to denote that $\rew^{i+1}(v)$ results from 
	$\rew^i(v)$ after choosing the atom $\alpha_i$ from the body of $\rew^i(v)$ at the beginning of the $i$-th rewriting step.
	
	Below, we describe a process that computes a new ruleset by rewriting the rules associated with the nodes of an EG.  
	\begin{definition}\label{definition:rule-rewriting}
		Let $G$ be an EG of a program P with a single leaf node.
		Let $\pi$ be a mapping associating each edge ${\varepsilon  \in \edges{G} \cup \{\diamond\}}$ with $\diamond$ denoting the empty edge, with a fresh predicate ${\pi(\epsilon)}$. 
		We denote by ${\rho(G,\pi)}$ the rules obtained from the rules associated with the nodes in $G$ after rewriting them as follows:  
		replace each $A(\vec{X})$ that is either 
		 \begin{compactitem}
			\item i. the $i$-th intensional atom in the body of $\rul{v}$ or the head atom of $\rul{u}$ and ${\varepsilon \defeq u \rightarrow_i v}$ in $\edges{G}$; or 
			\item ii. the head atom of $\rul{\kappa}$ with  $\kappa$ being the leaf node of $G$,
		\end{compactitem}
		by $A^*(\vec{X})$, where ${A^* = \pi(\varepsilon)}$ when (i) holds, or ${A^* = \pi(\diamond)}$ when (ii) holds.  
		
		For a node ${u \in \nodes{G}}$, we denote by $r_u^{\pi}$ the rule from $\rho(G,\pi)$ that results after rewriting $\rul{u}$.  
	\end{definition}
	
	From Definition~\ref{definition:rule-rewriting} we can see that the following holds: 
	\begin{corollary}\label{corollary:sharing-predicates}
		For an EG $G$ of a program P with a single leaf node, a mapping $\pi$ associating each edge ${\varepsilon \in \edges{G} \cup \{\diamond\}}$ 
		with a fresh predicate and two rules $\varrho_1$ and $\varrho_2$ from ${\rho(G,\pi)}$, we have: 
		the $i$-th body atom of $\varrho_1$ has the same predicate with the head atom of $\varrho_2$ only 
		if $\varrho_j$ is of the form ${r_{u_j}^{\pi}}$, for ${1 \leq j \leq 2}$ and ${u_2 \rightarrow_i u_1}$ is in $\edges{G}$.   
	\end{corollary}

	Let ${\Gamma = G_{\preceq v}}$ and $n$ be the depth of $\Gamma$. 
	Let $\pi$ be a mapping from edges to predicate as defined above. 
	Let $P^*$ be the rules in ${\rho(\Gamma,\pi)}$.
	Let $R$ be the predicate occurring in the head of $\rul{v}$ 
	and let ${\head{\rul{v}} = R(\vec{Y})}$. 
	Let ${R^* = \pi(\diamond)}$. 
	Let ${Q(\vec{Y}) \leftarrow R(\vec{Y})}$ and ${Q^*(\vec{Y}) \leftarrow R^*(\vec{Y})}$. 
	Let ${\mathcal{T}^{i}}$ be the tree computed at the end of the $i$-th iteration of ${\mathsf{XRewrite}(P^*,Q^*)}$.   
	
	Below, we establish the relationship between the $\rew{v}$ and the rewritings computed by $\mathsf{XRewrite}$.
	\begin{lemma}\label{lemma:correspondence-of-rewritings}
		For each branch ${\kappa^0 \xrightarrow{\epsilon_0} \dots \xrightarrow{\epsilon_{n}} \kappa^{n+1}}$ with ${\epsilon_i = (\beta_i,\varrho_i)}$ and 
		$\kappa^i$ is a node of depth $i$ in ${\mathcal{T}^{n+1}}$, 
		there exists a sequence 
		\begin{align}
			\rew^0(v) \xrightarrow{\alpha_0} \dots \xrightarrow{\alpha_{n}} \rew^{n+1}(v)
		\end{align}
		such that $\rew^{n+1}(v)$ equals $\qry{\kappa^{n+1}}$ modulo bijective variable renaming. 
	\end{lemma}
	The proof of Lemma~\ref{lemma:correspondence-of-rewritings} follows from:
	(i) ${Q(\vec{Y}) \leftarrow R(\vec{Y})}$ and ${Q^*(\vec{Y}) \leftarrow R^*(\vec{Y})}$, where ${R^* = \pi(\diamond)}$,  
	(ii) the correspondence between the rewriting steps (1)--(3) of $\mathsf{XRewrite}$ and the rewriting process of Definition~\ref{definition:EP-rewriting}
	and (iii) Corollary~\ref{corollary:sharing-predicates}. 	

	From Lemma~\ref{lemma:correspondence-of-rewritings} and since $\rew^{n+1}(v)$ includes only \EP-atoms, we have that:
	${\mathsf{XRewrite}(P^*,Q^*)}$ terminates after $n+1$ iterations. 
	Furthermore, since $\mathsf{XRewrite}$ terminates and due to its correctness we have: 
	$R^*(\vec{t}) \in \Step{n}{P^*, B}$ iff $\vec{t}$ is an answer to $\mathsf{XRewrite}(P^*,Q^*)$ on $B$. 
	
	Due to Corollary~\ref{corollary:sharing-predicates}, we also have: 
	
	\begin{lemma}\label{lemma:correspondence-of-facts}	
		For each base instance $B$, the following inductive property holds for each ${1 \leq i \leq n+1}$:
		\begin{compactitem}
			\item $\phi$. ${S(\vec{t}) \in u(B)}$, with $u$ being a node of depth $i$ in $\Gamma$ iff $S^*(\vec{t}) \in \Step{i}{P^*, B}$, where $S^*$ is the predicate of the head atom of $r_u^{\pi}$.
		\end{compactitem}
	\end{lemma}
	
	We now establish the correspondence between $v(B)$ and the answers to $\rew(v)$ on $B$.  
	From Lemma~\ref{lemma:correspondence-of-facts}, we have:
	$R^*(\vec{t}) \in \Step{n}{P^*, B}$ iff ${R(\vec{t}) \in v(B)}$.  
	Since $R^*(\vec{t}) \in \Step{n}{P^*, B}$ iff $\vec{t}$ is an answer to $\mathsf{XRewrite}(P^*,Q^*)$ on $B$, 
	it follows that ${R(\vec{t}) \in v(B)}$ iff $\vec{t}$ is an answer to $\mathsf{XRewrite}(P^*,Q^*)$ on $B$. 
	Since $\vec{t}$ is an answer to $\mathsf{XRewrite}(P^*,Q^*)$ on $B$ iff $\vec{t}$ is an answer to $\rew(v)$ on $B$
	and due to the above, 
	it follows that: $\vec{t}$ is an answer to $\rew(v)$ on $B$ iff ${R(\vec{t}) \in v(B)}$.
	The above completes the proof of Lemma~\ref{lemma:EP-rewriting}. 
\end{proof}

%% file: appendix/minDatalog.tex
\begin{lemma}\label{lemma:EG-equivalence}
	For each EG $G$ for a Datalog program $P$ and each base instance $B$ of $P$, ${G(B) = \mathsf{minDatalog}(G)(B)}$. 
\end{lemma}
\begin{proof}
	Let $G^i$ be the EG at the beginning of the $i$-th iteration of $\mathsf{minDatalog}(G)$ with ${G^0 = G}$. 
	We show that the following property holds for each ${i \leq 0}$:
	\begin{compactitem}
		\item $\phi$. ${G(B) = \mathsf{minDatalog}(G^i)(B)}$.
	\end{compactitem}
	
	For ${i=0}$, $\phi$ trivially holds, since ${G^0 = G}$.
	For ${i+1}$ and assuming that $\phi$ holds for ${i \leq 0}$, we have.
	Let $u$, $v$ be a pair of nodes in $\nodes{G^i}$, 
	such that (i) $v$'s depth is not less than $u$'s depth,
	(ii) the predicates of $\head{\rul{v}}$ and of $\head{\rul{u}}$ are the same and (iii) the EG-rewriting of $v$
	 is contained in the EG-rewriting of $u$. 
	In order to show that the inductive property for $\phi$, it suffices to show that for each node ${w \in \nodes{G^i}}$ 
	for which ${v \rightarrow_j w \in \edges{G^i}}$ holds for some $j$, and each base instance $B$ of $P$, $w(B)$ is the same both in ${G^i(B)}$ and in ${G^{i+1}(B)}$. 
	However, this holds since (i) for each ${v \rightarrow_j w \in \edges{G^i}}$, we have ${u \rightarrow_j w \in \edges{G^{i+1}}}$, (ii) ${R(\vec{t}) \in v(B)}$ 
	implies ${R(\vec{t}) \in u(B)}$ and ${G^i_{\preceq u} = G^{i+1}_{\preceq u}}$. The above shows that $\phi$ holds for ${i+1}$ and concludes the proof of 
	Lemma~\ref{lemma:EG-equivalence}. 
\end{proof}

\thmminimizeDatalog*

\begin{proof}
	\emph{Part I}. The proof follows from Lemma~\ref{lemma:EG-equivalence}.

	\emph{Part II}. 
	First, we can see that the following holds due to Definition~\ref{definition:minimizationDatalog}:
	\begin{corollary}\label{corollary:minimizeDatalog}
		For a TG $G$ of a Datalog program $P$, there exists no two nodes $u,v$ in $\mathsf{minDatalog}(G)$ satisfying the following: 
		(i) $u$ and $v$ define the same predicate\footnote{We say that a node $u$ in a TG defines a predicate $A$ if the predicate of $\head{u}$ is $A$.} $A$ and  
		(ii) ${\rew(u) \subseteq \rew(v)}$. 
	\end{corollary}
	
	The proof works by contradiction.
	Let $P$ be a Datalog program, $G$ be a TG for $P$ and ${\Gamma = \mathsf{minDatalog}(G)}$. Suppose by contradiction that there exists a TG $\Gamma'$ for $P$ with 
	${\nodes{\Gamma} > \nodes{\Gamma'}}$. From Lemma~\ref{lemma:EP-rewriting} we know that for each set of nodes ${u_1, \dots, u_m}$ from $\nodes{\Gamma}$ 
	defining a predicate $A$, 
	there exists a set of nodes ${u'_1,\dots, u'_n}$ from $\nodes{\Gamma'}$ defining also $A$, such that the following holds: 
	\begin{align}
		\rew(u_1) \cup \dots \cup \rew(u_m) \equiv \rew(u'_1) \cup \dots \cup \rew(u'_n) \label{eq:minimizeDatalog}
	\end{align} 
	
	Since ${\nodes{\Gamma} > \nodes{\Gamma'}}$, we know that there exist a set ${u_1,\dots, u_m}$ and a set ${u'_1,\dots, u'_n}$ so that ${m > n}$. 
	Since \eqref{eq:minimizeDatalog} holds, we know from \cite{10.1145/322217.322221} that the following hold:  
	\begin{compactitem}
		\item for each ${\rew(u_i)}$ with ${1 \leq i \leq m}$, there exists a ${\rew(u'_j)}$ with ${1 \leq j \leq n}$, 
		such that ${\rew(u_i) \subseteq \rew(u’_j)}$;
		\item for each ${\rew(u'_j)}$ with ${1 \leq j \leq n}$, there exists a ${\rew(u_\ell)}$ with ${1 \leq \ell \leq n}$, 
		such that ${\rew(u'_j) \subseteq \rew(u_\ell)}$. 
	\end{compactitem}
	
	Since ${m > n}$, it follows that there exist ${i_1, i_2}$ with ${1 \leq i_1, i_2 \leq m}$ 
	and an $\ell$ with ${1 \leq \ell \leq n}$, such that ${\rew(u_{i_1}) \subseteq \rew(u'_\ell)}$ and ${\rew(u_{i_2}) \subseteq \rew(u'_\ell)}$ hold.  
	Below, we show how we reach a contradiction. 
	We consider the following cases: 
	\begin{compactitem}
		\item there exists an $i_3$ with  ${1 \leq i_3 \leq m}$ and ${i_3 \neq i_1,i_2}$, such that ${\rew(u'_\ell) \subseteq \rew(u_{i_3})}$.
		From the above, it follows that ${\rew(u_{i_1}) \subseteq \rew(u_{i_3})}$ and ${\rew(u_{i_2}) \subseteq \rew(u_{i_3})}$ leading to a contradiction 
		due to Corollary~\ref{corollary:minimizeDatalog}. 
		
		\item ${\rew(u_{i_1}) \subseteq \rew(u'_\ell)}$. From the above, it follows that ${\rew(u_{i_2}) \subseteq \rew(u_{i_1})}$ leading again to a contradiction
		due to Corollary~\ref{corollary:minimizeDatalog}. 
	\end{compactitem}
	The above completes the proof of Theorem~\ref{theorem:minimizeDatalog}.  
\end{proof}

%% file: appendix/complexity_minDatalog.tex
\complexityMinimumDatalog*

\begin{proof}
\emph{Membership}.
We show that deciding wether $G$ is a TG of $P$ \emph{not} of minimum size is in NP.
By Definition~\ref{definition:minimizationDatalog} and Theorem~\ref{theorem:minimizeDatalog}, $G$ is a TG of $P$ not of minimum size iff there exists a pair of vertices $u$ and $v$ in $G$ satisfying the conditions in Definition~\ref{definition:minimizationDatalog} for which ${\rew(v)} \subseteq {\rew(u)}$ (remember that the last condition holds iff there exists a homomorphism from ${\rew(u)}$ to ${\rew(v)}$).
Hence, to \emph{disprove} that $G$ is a TG of $P$ of minimum size, it is sufficient to guess such nodes $u$ and $v$, guess the homomorphism from ${\rew(u)}$ to ${\rew(v)}$ (observe that the size of ${\rew(u)}$ and ${\rew(v)}$ is polynomial), then compute ${\rew(u)}$ and ${\rew(v)}$ (feasible in deterministic polynomial time), and then check that the guessed homomorphism is correct (feasible in deterministic polynomial time).
This procedure is feasible in NP.

\emph{Hardness}.
We show the co-NP-hardness of the problem by showing the NP-hardness of its complement.
The reduction is from the NP-complete problem of query containment in relational DBs:
given two CQs $Q_1(\vec{X})$ and $Q_2(\vec{X})$ for a relational DB, decide whether $Q_1(\vec{X})\subseteq Q_2(\vec{X})$.
Let the queries be $Q_1(\vec{X}) \leftarrow a_{i_1}(\vec{X}_{i_1}),\dots,a_{i_n}(\vec{X}_{i_n})$ and $Q_2(\vec{X}) \leftarrow a_{j_1}(\vec{X}_{j_1}),\dots,a_{j_m}(\vec{X}_{j_m})$.

We now describe the reduction.
Consider the following program $P$ and TG $G$.

The rules of $P$ are obtained as follows.
Let $D = \{a_{k_1}, \dots, a_{k_\ell}\}$ be the set of all the \emph{distinct} predicates from $\{a_{i_1},\dots,a_{i_n}\}$.
For each of the predicates $a_{k_t}$ in $D$, there is a rule $a_{k_t}(\vec{X}_{k_t}) \rightarrow A_{k_t}(\vec{X}_{k_t})$ in $P$.
In $P$ there are also the rules $A_{i_1}(\vec{X}_{i_1}),\dots,A_{i_n}(\vec{X}_{i_n}) \rightarrow Q(\vec{X})$ and $a_{j_1}(\vec{X}_{j_1}),\dots,a_{j_m}(\vec{X}_{j_m}) \rightarrow Q(\vec{X})$.

The TG $G$ is as follows.
There is a node $v_{k_t}$ associated with each of the rules $a_{k_t}(\vec{X}_{k_t}) \rightarrow A_{k_t}(\vec{X}_{k_t})$;
there is a node $v$ associated with the rule $A_{i_1}(\vec{X}_{i_1}),\dots,A_{i_n}(\vec{X}_{i_n}) \rightarrow Q(\vec{X})$;
and there is a node $u$ associated with the rule $a_{j_1}(\vec{X}_{j_1}),\dots,a_{j_m}(\vec{X}_{j_m}) \rightarrow Q(\vec{X})$.
The edges of $G$ are:
for each $1\leq s\leq n$, there is an edge labelled $s$ to node $v$ from the node $v_{k_t}$ such that the predicate of $\head{\rul{v_{k_t}}}$ is $A_{i_s}$.

We show the $G$ is a TG of minimum size for $P$ iff $Q_1(\vec{X})\subseteq Q_2(\vec{X})$.

First, observe that the predicates of the rules associated with nodes $v_{k_t}$ are all distinct, and they differ from the predicate of the heads of the rules associated with $u$ and $v$. Hence none of the nodes $v_{k_t}$ can be removed from $G$ in the minimization process.
Nodes $u$ and $v$ are the only nodes in $G$ associated with rules with the same head predicate.
The depth of $u$ is 0, while the depth of $v$ is 1.
Hence, $v$ is the only node that can be removed in the minimization process.
Therefore, $G$ is not of minimum size iff $v$ can be removed.
The node $v$ can be removed iff ${\rew(v)} \subseteq {\rew(u)}$, and hence, by the definition of $P$, iff $Q_1(\vec{X})\subseteq Q_2(\vec{X})$.
\end{proof} 

%% file: appendix/TGmat.tex
\thmTGmat*

\begin{proof}
	We first show that 
	
	\begin{claim}\label{claim:EGexec}
		For each node ${v \in \nodes{G}}$ and each instance $I$, we have
		\begin{align}
			v(B,I) = v(B) \setminus I	
		\end{align}
	\end{claim}
	
	\begin{proof}
		Let ${A(\vec{X})}$ be the head atom of $\rul{v}$ and let
		${Q(\vec{Y}) \leftarrow \bigwedge \nolimits_{i = 1}^n f_i}$ be the EG-rewriting of $v$.
		
		Recall from Lemma~\ref{lemma:EP-rewriting} that for each base instance of $B$ of $P$ we have: 
		$v(B)$ includes exactly a fact ${A(\vec{t})}$ for each answer $\vec{t}$ to the EG-rewriting of $v$ on $B$.
		
		Now consider any $m \geq 1$ atoms ${f_{i_1}, \dots, f_{i_m}}$ from the body of $Q$ whose variables include all variables in ${\vec{Y}}$. 
		Consider also the query ${Q'(\vec{Y}) \leftarrow f_{i_1} \wedge \dots \wedge f_{i_m}}$. From \cite{Cali:2013:TIC:2591248.2591252}, 
		it follows that $Q$ is contained in $Q’$, i.e., for each base instance $B$, each answer $\vec{t}$ to $Q$ on $B$ is an answer to $Q’$ on $B$. 
		From the above, we have: each $\vec{t}$, for which ${A(\vec{t}) \in v(B)}$ holds, 
		is also an answer to ${Q'(\vec{Y}) \leftarrow f_{i_1} \wedge \dots \wedge f_{i_m}}$ on $B$. 
		We refer to this conclusion as $(\ast)$.      
		
		Since step (2) of Definition~\ref{definition:EGexec} considers each homomorphism $h$ for which (i) 
		${h(\vec{X})}$ is an answer to $Q'$ on $B$ and (ii) ${A(h(\vec{X})) \not \in I}$ and due to $(\ast)$, it follows that Claim~\ref{claim:EGexec} holds. 	  
	\end{proof}
	
	Let $I^k$ be the instance computed the beginning of the $k$-th iteration of the steps in 
	lines~\ref{algorithm:while:begin}--\ref{algorithm:while:end} of Algorithm~\ref{alg:online}. 
	Then, using Claim~\ref{claim:EGexec}, Theorem~\ref{lemma:kboundedness:chase-equivalence} 
	and Lemma~\ref{lemma:EG-equivalence}, we can easily show that for each ${k \geq 0}$, the following property holds:
	\begin{compactitem}
		\item $\phi$. ${I^k = \Step{k}{P, B}}$
	\end{compactitem} 
	
	The above concludes the proof of Theorem~\ref{theorem:TGmat}.	
\end{proof}

%% file: appendix/examples.tex
\section{Additional Examples}

\begin{example}\label{example:guided-chase}
	We show how reasoning over the TG $G_1$ from Figure~\ref{figure:trigger-graphs} proceeds
	for the base instance ${B = \{\rs(c_1,c_2)\}}$.
	
	Reasoning starts from the root nodes $u_1$ and $u_2$, which are associated with the rules $r_1$ and $r_4$, respectively.
	Since there exists a homomorphism ${h=\{X \mapsto c_1, Y \mapsto c_2\}}$ from $\body{r_1}$ into $B$
	and from $\body{r_4}$ into $B$, we have
	\begin{align}
		u_1(\{f_1\}) &= \{ \Rp(c_1,c_2) \} \label{eq:u_1_f_1} \\
		u_2(\{f_1\}) &= \{T(c_2,c_1,\lnu_1)\} \label{eq:u_2_f_1}
	\end{align}
	where $\lnu_1$ is a null. 
	Then, since there exists an edge from $u_1$ to $u_3$ and since $u_3$ is associated with $r_2$,
	we compute all homomorphisms from $\body{r_2}$ into $u_1(B)$.
	Since there exists a homomorphism ${h=\{X \mapsto c_1, Y \mapsto c_2\}}$
	from $\body{r_2}$ into $u_1(B)$, we have 
	\begin{align}
		u_3(\{f_1\}) &= \{T(c_2,c_1,c_2)\} \label{eq:u_3_f_1}
	\end{align}
	Since there is no other node, reasoning stops. 
\end{example}

\begin{example}
	We demonstrate the notion of preserving homomorphisms introduced in Definition~\ref{definition:preserving}. 

	Consider the facts ${f_1 = \rs(c_1,c_2)}$ and ${f_2 = \rs(c_3,c_3)}$ from the set $\mathcal{H}(P_1)$.
	By applying Definition~\ref{definition:guided-chase} for the base instance ${\{f_1\}}$, we have
	${u_1(\{f_1\})}$, ${u_2(\{f_1\})}$ and ${u_3(\{f_1\})}$ as in 
	\eqref{eq:u_1_f_1}, \eqref{eq:u_2_f_1} and \eqref{eq:u_3_f_1}. 
	Similarly, by applying Definition~\ref{definition:guided-chase} for the base instance ${\{f_2\}}$, we have
	\begin{align}
		u_1(\{f_2\}) &= \{ \Rp(c_3,c_3) \}\\
		u_2(\{f_2\}) &= \{T(c_3,c_3,\lnu_2)\}\\
		u_3(\{f_2\}) &= \{T(c_3,c_3,c_3)\}
	\end{align}
	Above, $\lnu_2$ is a null.  
	We can see that there exists a preserving homomorphism from ${u_2(\{f_1\})}$ into ${u_3(\{f_1\})}$ mapping $\lnu_1$ to $c_2$, 
	since the null $\lnu_1$ is not shared among the facts occurring in the instances associated with $u_2$ and $u_2$. 
	For the same reason, there exists a preserving homomorphism from ${u_2(\{f_2\})}$ into ${u_3(\{f_2\})}$ mapping $\lnu_2$ to $c_3$.
	Hence according to Lemma~\ref{lemma:linear:redundant}, there exists a preserving homomorphism from  
	$u_2(B)$ into $u_3(B)$ for each base instance $B$. 
\end{example}

\begin{example}
	We demonstrate the computation of EG-rewritings introduced in Definition~\ref{definition:EP-rewriting}. 
	
	Consider the rules 
	\begin{align*}
		\rs(X_1,Y_1,Z_1) &\rightarrow \T(X_1,X_1,Y_1) \tag{$r_{10}$} \\
		\T(X_2,Y_2,Z_2) &\rightarrow \Rp(Y_2,Z_2) \tag{$r_{11}$}
	\end{align*}
	where $\rs$ is the only extensional predicate. 
	Consider now an EG having nodes $u_1$ and $u_2$, where $u_i$ is associated with $r_i$ for each ${1 \leq i \leq 2}$, 
	and the edge ${u_1 \rightarrow_1 u_2}$.  
	
	To compute the EG-rewriting ${\rew(u_2)}$ of $u_2$ we first form the query 
	\begin{align}
		Q(Y_2,Z_2) \leftarrow \Rp(Y_2,Z_2)												\label{eq:query0}
	\end{align}
	and associate the atom ${\Rp(Y_2,Z_2)}$ with $u_2$. 
	The following steps take place in the first iteration of the rewriting algorithm. 
	First, since ${\Rp(Y_2,Z_2)}$ is the only intensional atom in the query 
	we have ${\alpha = \Rp(Y_2,Z_2)}$. Then, according to step (ii) and since the node $u_2$ is associated with $\Rp(Y_2,Z_2)$,  
	we compute the MGU $\theta_1$ of the set ${\{\head{u_2},\Rp(Y_2,Z_2)\}}$.
	We have $\theta_1 = \{Y_2 \rightarrow Y_2, Z_2 \rightarrow Z_2 \}$, since ${\head{u_2} = \Rp(Y_2,Z_2)}$.
	By applying the step (iii), the query in \eqref{eq:query0} becomes 
	\begin{align}
		Q(Y_2,Z_2) \leftarrow \T(X_2,Y_2,Z_2)											\label{eq:query1}
	\end{align}
	In step (iv) we associate the fact ${\T(X_2,Y_2,Z_2)}$ with node $u_1$ due to the edge 
	${u_1 \rightarrow_1 u_2}$. 
	
	In the second iteration of the rewriting algorithm, we have ${\alpha = \T(X_2,Y_2,Z_2)}$. 
	Since the fact ${\T(X_2,Y_2,Z_2)}$ is associated with node $u_1$, 
	in step (ii) we compute the MGU $\theta_2$ of the set ${\{\head{u_1},\T(X_2,Y_2,Z_2)\}}$. 
	We have $\theta_2 = \{X_1 \rightarrow Y_2, X_2 \rightarrow Y_2,  Y_1 \rightarrow Z_2\}$. 
	In step (iii) we replace ${\alpha = \T(X_2,Y_2,Z_2)}$ in \eqref{eq:query1} with ${\body{u_1} = \rs(X_1,Y_1,Z_1)}$
	and apply $\theta_2$ to the resulting query. The query in \eqref{eq:query1} becomes
	\begin{align}
		Q(Y_2,Z_2) \leftarrow r(Y_2,Z_2,Z_1)												\label{eq:query2}
	\end{align}
	Since there is no incoming edge to $u_1$, we associate no node to the fact ${r(Y_2,Z_2,Z_1)	}$.
	The algorithm then stops, since there is no extensional fact in \eqref{eq:query2}.
	The EG-rewriting of $u_2$ is the query shown in \eqref{eq:query2}.
\end{example}

\input{figures/full-graph}

\begin{example}\label{example:fullEGs}
	We demonstrate the notion of compatible nodes introduced in Definition~\ref{eq:compatible-nodes}, 
	as well as the procedure for computing instance-dependent TGs from Section~\ref{section:trigger-graph}. 

	Consider the program $P_3$
	\begin{align}
		\as(X) &\rightarrow \A(X) 								\subtag{$r_{12}$} \\
		\rs(X,Y) &\rightarrow \Rp(X,Y) 							\subtag{$r_{13}$} \\
		\Rp(X,Y) \wedge \A(Y) &\rightarrow \A(X) 				\subtag{$r_{14}$} \\
		\Rp(X,Y) \wedge \Rp(Y,Z) &\rightarrow \A(X) 				\subtag{$r_{15}$}
	\end{align}
	where $\as$ and $\rs$ are extensional predicates. 
	Figure~\ref{figure:full-graph} shows part of the graph computed up to level 3.
	Next to each node, we show the rule associated with it.
	For example, node $u_1$ is associated with rule $r_{13}$ and node $u_2$ is associated with rule $r_{12}$.
	
	When $k=1$, $G^1$ includes two nodes, one associated with rule $r_{12}$ ($u_2$) and one associated with rule $r_{13}$ ($u_1$). 
	When ${k=2}$, $r_{14}$ has only one $2$-compatible combination of nodes. That is ${(u_1, u_2)}$.
	Hence, the technique will add one fresh node $u_3$, associated with $r_{14}$ and will add the edges 
	${u_1 \rightarrow_1 u_3}$ and ${u_2 \rightarrow_2 u_3}$. 
	The $2$-compatible combination of nodes for $r_{15}$ is ${(u_2, u_2)}$.
	Hence, the technique will add one fresh node $u_4$, associated with $r_15$ and will add the edges 
	${u_2 \rightarrow_1 u_4}$ and ${u_2 \rightarrow_2 u_4}$. 
	
	When ${k=3}$, $r_{14}$ has the following $3$-compatible combinations of nodes:
	${(u_1, u_3)}$ and ${(u_1, u_4)}$. 
	For each such $3$-compatible combinations of nodes, the algorithm adds a fresh node and associates it with $r_{14}$. 
	For ${k=3}$, the $3$-compatible combinations of nodes for $r_{15}$ are: 
	${(u_2, u_3)}$, ${(u_2, u_4)}$, ${(u_3, u_2)}$, ${(u_4, u_2)}$,   
	${(u_3, u_4)}$, ${(u_4, u_3)}$, ${(u_3, u_3)}$, ${(u_4, u_4)}$.
	Again, for each such combination of nodes, the algorithm adds a fresh node and associates it with $r_{15}$. 
\end{example}

%% file: figures/full-graph.tex
\begin{figure}[tb]
\begin{center}
\scalebox{0.8}{
	\newcommand{\xsep}{3}
	\newcommand{\ysep}{1.5}
	\begin{tikzpicture}[roundnode/.style={circle, fill=black, inner sep=0pt, minimum size=1.2mm}]
	\node[roundnode, fill=black, label={left}:{$u_1\backslash r_{13}$}] at (1*\xsep, 2.5*\ysep) (u1) {};
	\node[roundnode, fill=black, label={left}:{$u_2\backslash  r_{12}$}] at (1*\xsep, 3*\ysep) (u2) {};
	
	\node[roundnode, fill=black, label={above}:{$u_3\backslash  r_{14}$}] at (2*\xsep, 3*\ysep) (u3) {};
	\node[roundnode, fill=black, label={below}:{$u_4\backslash  r_{15}$}] at (2*\xsep, 2*\ysep) (u4) {};
	
	\node[roundnode, fill=black, label={right}:{$u_5\backslash  r_{14}$}] at (3*\xsep, 3*\ysep) (u5) {};
	\node[roundnode, fill=black, label={right}:{$u_6\backslash  r_{14}$}] at (3*\xsep, 2*\ysep) (u6) {};
	
	\path[->] (u1) edge node [fill=white,pos=0.5] {\small{1}} (u3);
	\path[->] (u1) edge[bend right = 12] node [fill=white,pos=0.4] {\small{2}}  (u4);
	
	\path[->] (u1) edge [bend left = 12] node [fill=white,pos=0.7] {\small{1}} (u4);
	
	\path[->] (u1) edge[bend right = 4] node [fill=white,pos=0.8] {\small{1}}  (u5);
	\path[->] (u1) edge[bend left = 4] node [fill=white,pos=0.8] {\small{1}} (u6);
	
	\path[->] (u2) edge node [fill=white,pos=0.3] {\small{2}} (u3);
	
	\path[->] (u4) edge node [fill=white,pos=0.3] {\small{2}} (u6);
	\path[->] (u3) edge node [fill=white,pos=0.3] {\small{2}} (u5);
	
	\end{tikzpicture}
}
\end{center}
\caption{Part of the graph from Example~\ref{example:fullEGs}.}
\label{figure:full-graph}
\end{figure}
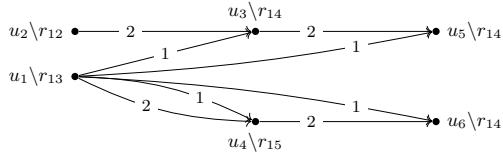